\RequirePackage{fix-cm}
\documentclass[twocolumn]{svjour3}       
\smartqed  

\makeatletter
\expandafter\let\csname opt@amsmath.sty\endcsname\relax
\AtBeginDocument{
	\mathindent=15pt 
	\@mathmargin\@centering} 
\makeatother

\usepackage{graphicx}
\usepackage{booktabs} 

\usepackage{amsmath}
\usepackage[ruled, linesnumbered ]{algorithm2e}
\usepackage{subfigure}
\usepackage{graphicx}
\usepackage{diagbox}
\usepackage{caption}

\usepackage{multirow}

\usepackage{bm}
\usepackage{amsmath}
\usepackage{amssymb}

\usepackage{multicol}
\usepackage{multirow}

\usepackage{lipsum}

\usepackage{mathtools}

\usepackage{marvosym}

\usepackage{xcolor}

\newcommand{\M}{\mathcal{M}}  

\newcommand{\D}{d}  
\newcommand{\Domain}{\mathcal{D}}
\newcommand{\Range}{\mathcal{R}}
\newcommand{\eps}{\varepsilon}

\begin{document}

\title{Fast-adapting and Privacy-preserving Federated  Recommender System}


\author{Qinyong Wang\textsuperscript{1}  \and Hongzhi Yin\textsuperscript{1$\ast$} \and Tong Chen\textsuperscript{1} \and Junliang Yu\textsuperscript{1} \and Alexander Zhou\textsuperscript{2} \and Xiangliang Zhang\textsuperscript{3} \thanks{$\ast$ Corresponding author: Hongzhi Yin.} }


\institute{
	Qinyong Wang  \at
	\email{qinyong.wang@uq.edu.au}
	\and
	Hongzhi Yin  (\Letter) \at
	\email{h.yin1@uq.edu.au}
	\and
	Tong Chen \at
	\email{tong.chen@uq.edu.au}
	\and
	Junliang Yu \at
	\email{jl.yu@uq.edu.au}
	\and
	Alexander Zhou \at
	\email{atzhou@cse.ust.hk}
	\and
	Xiangliang Zhang \at
	\email{xiangliang.zhang@kaust.edu.sa} \at
	\begin{description}
		\item[\textsuperscript{1}] School of Information Technology
		and Electrical Engineering, The University of Queensland, Australia
		\item[\textsuperscript{2}] Hong Kong University of	Science and Technology, Hong Kong
		\item[\textsuperscript{3}] King Abdullah University of Science and Technology, Saudi Arabia
	\end{description}	
}

\date{Received: date / Accepted: date}

\maketitle

\begin{abstract}
In the mobile Internet era, recommender systems have become an irreplaceable tool to help users discover useful items, thus alleviating the information overload problem.
Recent research on deep neural network (DNN)-based recommender systems have made significant progress in improving prediction accuracy, largely attributed to the widely accessible large-scale user data. Such data is commonly collected from users' personal devices, and then centrally stored in the cloud server to facilitate model training. However, with the rising public concerns on user privacy leakage in online platforms, online users are becoming increasingly anxious over abuses of user privacy. Therefore, it is urgent and beneficial to develop a recommender system that can achieve both high prediction accuracy and strong privacy protection.

To this end, we propose a DNN-based recommendation model called PrivRec running on the decentralized federated learning (FL) environment, which ensures that a user's data is fully retained on her/his personal device while contributing to training an accurate model. On the other hand, to better embrace the data heterogeneity (e.g., users' data vary in scale and quality significantly) in FL, we innovatively introduce a first-order meta-learning method that enables fast on-device personalization with only a few data points.
Furthermore, to defend against potential malicious participants that pose serious security threat to other users, we further develop a user-level differentially private model, namely DP-PrivRec, so attackers are unable to identify any arbitrary user from the trained model. To compensate for the loss by adding noise during model updates, we introduce a two-stage training approach.
Finally, we conduct extensive experiments on two large-scale datasets in a simulated FL environment, and the results validate the superiority of both PrivRec and DP-PrivRec. 
\end{abstract}

\section{Introduction}

\subsection{Background}
With the immense popularity of mobile phones, users can gain access to a large  number of online content and services with only one click, such as news, e-commerce, movies and music. However, it has also become difficult for users to accurately find information relevant to their interests. Therefore, recommender systems have become an essential tool for alleviating such information overload by automatically generating a recommendation list based on a user's preference.

Meanwhile, the past decade has seen the enormous advancement of deep neural network (DNN)-based recommender systems~\cite{zhang2019deep}, which are able to achieve superior performance in terms of many aspects compared to conventional models.
On the other hand, it is well-known that the overwhelming performance of DNN-based models is largely attributed to the increased access to large-scale training data. Such recommender systems are usually deployed on a cloud server to acquire the sufficient resource for data storage and efficient training, which then provides services to connected users in an on-demand manner. This means that they require to collect users' behavioral data (e.g., browsing history) and then centrally store it in the service providers' central servers. 
Unfortunately, this centralized recommendation paradigm inevitably leads to increasing concerns on user privacy as user data stored on the server might be accidentally leaked or misused.
Even though the data sent to the server can be anonymized, users' information can still be identified when linked with other data sources~\cite{fung2010privacy}.
\subsection{Motivations and Proposals}

Therefore, it is important and desirable to offer a privacy-by-design solution where the recommendation service providers do not need to access personal data, while still building robust and accurate recommendation models. Over the past decade, the capacities of storage and computation increase dramatically on personal devices (e.g., smart phones and laptops), making it possible to distribute the resource-intensive model training process from the server to edge devices owned by users.
In this regard, federated learning (FL)~\cite{yang2019federated,fung2010privacy} has emerged as a popular framework in the development of privacy-preserving machine learning systems. Training an FL model consists of several key steps.
First, the server selects a batch of available devices and sends them the current global model. Then, based on locally stored user data, each device computes the gradients w.r.t. an objective function of the model. Finally, the central system aggregates the gradients from different devices in order to update the global model. These training steps are iterated until the global model converges. Each user's data is fully kept on her/his personal device in the FL setting  and thus they retain control over his/her own data. 

Motivated by the idea of FL, there have been some attempts~\cite{chen2018federated,jalalirad2019simple,ribero2020federating} to facilitate privacy-preserving recommendation  but they ignore two important problems in real-world scenarios, which are \textit{data heterogeneity} and \textit{malicious participants}.
In this paper, to address both challenges in a unified way, we propose two DNN-based recommender systems that smoothly run in the FL setting, namely \textbf{PrivRec} and the more secure \textbf{DP-PrivRec} enhanced by differential privacy. In what follows, we elaborate on these two challenges and our corresponding solutions.

\textbf{Data Heterogeneity.}
It is common that users have highly different personal preferences when consuming online products. Hence, the data heterogeneity problem arises in FL-based recommenders where the local data generated by different users vary significantly in both distribution and scale. This problem will be exacerbated for recommendation models due to two reasons.
First, frequent users have far more interaction records with items than other inactive (e.g., new or cold-start) users. Therefore, in a typical FL setting, straightforwardly minimizing the average loss for all users will mostly advantage the recommendation performance for active users while the inactive users' preferences can be hardly estimated.
On this occasion, the ultimate recommendation model is clearly unfair for a huge number of users due to biased service quality.
Second, the difficulty in achieving optimal personalization also arises with data heterogeneity. 
 Existing FL-based recommendation paradigms are only designed to let all devices contribute to a global model, hence the final model is only good on average recommendation accuracy. Since they do not account for the heterogeneity of data distribution among users, there is no guarantee that the global model is fully customized for every user, impeding the delivery of personalized recommendations.

In light of this, we resort to the notion of meta-learning~\cite{vanschoren2018meta} as a natural solution to address the data heterogeneity problem in the FL setting. 
Meta learning was originally proposed to quickly adapt the global model to a new task (a user in our case) using only a few data points. In a similar vein, some methods  ~\cite{chen2018federated,fallah2020personalized} utilize the recently proposed Model-Agnostic Meta-Learning (MAML)~\cite{finn2017model} to enable an FL recommender to achieve fast on-device personalization with  only a small amount of user-generated data.
However, inheriting the drawbacks of MAML, these methods incur huge computational cost when computing the second-order gradients, which is unbearable for the resource-constrained edge devices. Furthermore, they require to divide the local data into support and query sets for the two-stage training, which is impractical for users with a highly limited number of interaction records.
Hence, in this paper, we propose to use only the approximated first-order gradient for meta-learning (i.e., REPTILE~\cite{nichol2018first}), thus reducing computational burden while maintaining a comparable performance. Also no data splitting is required in our approach, making the model a better fit for both active and inactive users and easier to run in the FL environment.

We term this novel DNN-based FL recommender system coupled with the first-order meta-learning as PrivRec.

\textbf{Malicious Participants.} We show that PrivRec can effectively prevent it from sharing each users' sensitive data with the central server while training an accurate model. However, in an open Internet environment, it is insufficient to protect sensitive user data by simply decoupling the model training process from the need for direct access to the raw user data. More specifically, the unique settings of FL would increase the risk of privacy leakage by unintentionally allowing malicious clients to participate in the training. 

Sharing final trained models with other parties would usually incur three types of well-studied privacy attacks: membership inference, parameter inference and model inversion. 
Parameter inference attacks try to recover the model parameters or the hyper-parameters and model architecture. In the FL environment, these information is designed to be shared among the participants.
Meanwhile, model inversion attackers leverage auxiliary knowledge to
construct an inversion model which can reconstruct the original
input sample from the prediction scores with high accuracy. However, a malicious participant needs to know the auxiliary information about a particular user to start  model inversion attack, but in the  FL environment, a user's information never leaves his/her own device.
Therefore, we study how to defend the more feasible membership attack in FL, which tries to determine whether a given data point (i.e., a user's behavior footprints in our case) is used as part of the training set.

In~\cite{melis2019exploiting}, it is shown that a curious client in FL can infer not only membership, but also properties that characterize subsets of the data (e.g., sensitive user attributes).
Nasr \textit{et.al}~\cite{nasr2018comprehensive} comprehensively
explore membership inference on gradient parameters, including an analysis in an FL setting.
Orekondy \textit{et.al}~\cite{orekondy2018gradient} show that clients can be identified in an FL setting by model updates alone. 

\begin{figure}[!t]
	\includegraphics[scale = 0.47]{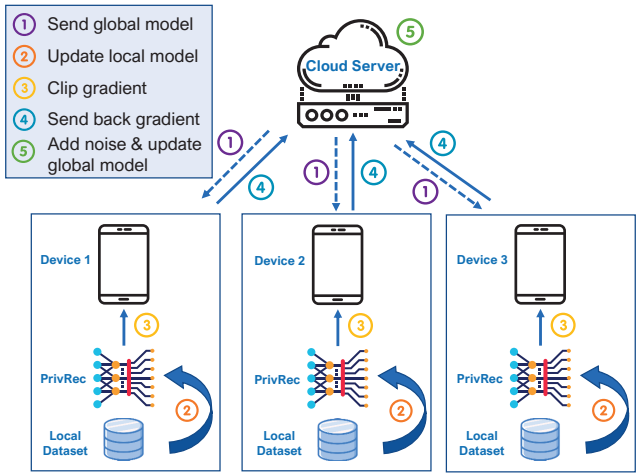}
	\caption{\textbf{An overview of our proposed DP-PrivRec. Numbers in the figure represent the orders in the pipeline we train DP-PrivRec. Note that to perturb the model parameters of PrivRec to  defend from adversary, additional and important steps are step 2 and step 5.}}
	\label{fig:net}
\end{figure}

Therefore, it is necessary to take further defensive measures so that the participating adversaries are unable to infer from the trained model whether a client has joined the decentralized training (i.e., membership inference). We present DP-PrivRec, an extension of PrivRec, to strengthen our FL-based recommender with differential privacy (DP)~\cite{abadi2016deep,dwork2008differential,dwork2006our,zhang2021graph}. DP is a well-established  method to defend against such attacks, whose basic idea is adding additional noise to ensure that the published information does not vary much whether one individual is present or not. In this way, the attackers cannot infer the private information of any user with high confidence, as no user significantly affects the final output.
The pipeline of running our proposed DP-PrivRec is briefly shown in Figure~\ref{fig:net}. Instead of only protecting a single data point like in conventional DP mechanisms, DP-PrivRec enforces differential privacy at the entire user level, which is more suited to FL-based recommenders as each user's published information is actually their gradient computed with the entire local dataset.
To achieve this, we employ the Gaussian mechanism~\cite{dwork2008differential} to perturb the gradients which are sent from the FL clients to the central server, so that the resulting recommender model satisfies a given privacy budget. 

However, it is inevitable that the recommendation performance would be damaged after noise is introduced in gradients. To address this problem, we enhance the item representation learning by developing a two-stage FL training method to compensate for the performance loss by DP. 
In the first stage, we only focus on learning item representations using  self-supervised learning (SSL) and without any interference from DP.
This is based on the observation that we could relax the level of privacy protection when user information is exclusive, which enhances the item representation learning.
In the second stage, we follow the training procedures of DP-PrivRec and optimize the model parameters by a downstream task, namely, modelling the user-item interactions.
We employ the well-trained item embeddings from the last stage as initialized values, and use the labeled user-item interactions to learn user embeddings and fine-tune item embeddings.


\subsection{Contributions}

We highlight our contributions below:
\begin{itemize}
	\item We develop a privacy-preserving recommendation model called PrivRec based on FL. Apart from preventing users from sharing their own data for model training, we propose an efficient and practical meta-learning approach to enable PrivRec to quickly adapt to inactive users, alleviating the critical data heterogeneity problem for existing FL-based recommenders.
	\item To defend against the malicious participants who may conduct membership inference attacks on the trained model, we propose DP-PrivRec with a user-level differential privacy mechanism. We also further introduce a two-stage training approach to compensate for the performance loss due to DP. 
	\item We perform extensive experiments to validate the effectiveness of our proposed models and results show that PrivRec and DP-PrivRec achieve superior performance over the state-of-the-art baselines.
\end{itemize}

The rest of this paper is organized as follows. Section~\ref{st:method} details our proposed methodologies, Section~\ref{se:exp} introduces the experimental settings, Section~\ref{st:result} presents the experimental results, Section~\ref{st:related} reviews related work and  finally  Section~\ref{st:conclude} concludes this paper.

\section{Methodologies}
\label{st:method}
In this section, we describe our proposed  PrivRec and DP-PrivRec models in detail.

Before we proceed, we list in Table~\ref{tb:notations} some important notations used throughout the paper.

\begin{table}[h]
	\centering
\caption{\textbf{Some important notations used in the paper.}}
	\begin{tabular}{|c|l|}
		 \hline
		\textbf{Notations}& \textbf{Description} \\ \hline
	${u}_i$, ${v}_j$	& user $i$, item $j$ \\ \hline
	$\theta_0$	& initial model parameters \\ \hline
	$\theta_\tau$ & model parameters for client $\tau$ \\ \hline
	$\theta_0$	& initial model parameters \\ \hline
	$E_1$,	$E_2$	& number of server and local training rounds \\ \hline
	$\alpha_1$,$\alpha_2$	&  learning rates for local and server updates \\ \hline
	$M$ & number of clients in each training round \\ \hline
	$S$ & clipping bound for DP \\ \hline
	$z$ & noise scale in DP \\ \hline
    $\epsilon$, $\delta$  & DP parameters \\ \hline
	
	\end{tabular}
\label{tb:notations}
\end{table}

\subsection{PrivRec in the Federated Learning Setting}

The federated learning (FL) framework~\cite{yang2019federated,bonawitz2019towards} aims to train a shared global model with a central server from decentralized data spanning over a large number of different clients.
Examples of such clients are smartphones, wearable devices or  medical institutions.
Most machine learning algorithms designed for FL solve the optimization problem  in multiple training rounds, where at each round the central server sends the current model to a fraction of the users and those users update the model w.r.t. their own loss functions.
Then, these users return their updated models to the central server, which combines the received models to update the global model and sends the updated model to another fraction of the users for the next round of training.
In this way, a model that solves the optimization problem is obtained and the resulted optimal model performs well over all users on average.

The advantage of FL that users' data can safely reside on their own devices makes it an attractive and popular alternative for developing privacy-preserving machine learning models, and it is desirable to extend a recommender system to the FL setting by simply treating each user as a client device. Note that we interchangeably use the term ``client'' and ``user'' in this paper.

On one hand, the deep neural network (DNN) techniques dominate  recommender systems such as~\cite{he2017neuralCF,yin2019social}, directly deploying them in the FL setting may still incur serious privacy disclosure.

When we train a user embedding matrix to keep the user preferences, its entries are the unique user  identification. Therefore, directly representing users and sharing this embedding matrix among the devices could be risky. Furthermore, it would be unfriendly for new/cold-start users who cannot be effectively represented in a fixed-size embedding matrix.  To avoid the dependency of user identifications for modelling, we instead take only user related features (e.g., age and occupation) (e.g., genre and textual description) as the input. For item representation, we concatenate all the related features, including the ID.

Specifically, for the recommendation model, we adopt the Deep Structured Semantic Models (DSSM)~\cite{huang2013learning}, a widely used information retrieval and recommendation network structure. In this model, the latent representations of user and item are not interacted until the last MLP layers, and the benefit of doing so is that we can separately train user and item representations. 

%

In the output layer, we compare the score with the actual user preference and minimize the loss. The user preferences could be explicit ratings or implicit feedbacks.

This  process could be formulated as:
\begin{equation}
\label{eq:mlp}
\begin{array}{l}
\mathbf{u}_{i} = f_u({u}_i), \mathbf{v}_{j} = f_v({v}_j),\\

{x_{0}=\left[\mathbf{u}_i , \mathbf{v}_{j}\right]}, \\
{x_{1}=Relu\left(W_{1}^{\top} x_{0}+b_{1}\right)}, \\
\quad \quad \quad  {\vdots} \\
{x_{N}=Relu\left(W_{N}^{\top} x_{N-1}+b_{N}\right)}, \\
{\hat{Y}_{i j}=f\left(W_{o}^{\top} x_{N}+b_{o}\right)},
\end{array}
\end{equation}
where ${u}_i$ and ${v}_j$ are the input representations for user $i$ and item $j$, $f_u$ and $f_v$ are deep networks encoding and transforming user and item respectively, $[.]$ is the concatenation operation, $W_{i}$ and $b_{i}$ are the weight and bias for the $i$-th MLP layer and $W_{o}$ and $b_{o}$ are the weight and bias for the output layer,
$\hat{Y}_{i j}$ is user $i$'s estimated preference for item $j$,  $Relu(.)$ denotes the rectified linear unit activation function for the hidden layers and $f(.)$ is the activation function of the output layer, which depends on how user preferences are modeled. If the target is implicit feedback, then the Sigmoid function could be used, and if the target is a rating, then a linear function is considered.

Take the Sigmoid activation function as an example,  the cross-entropy function is usually used to measure the loss:
\begin{equation}
\small
\mathcal{L}_\mathrm{DSSM} = \sum_{(i, j) \in Y^+\cup Y^{-}} Y_{i j} \log \hat{Y}_{i j}+\left(1-Y_{i j}\right) \log \left(1-\hat{Y}_{i j}\right),
\label{eq:dssm}
\end{equation}
where $Y$ and $Y^{-}$ are the sets of items that this user is interested or not, $Y_{i j} \in \{0,1\}$ is the ground-truth value.

\subsection{PrivRec with On-device Personalization}

It is known that FL suffers from the presence of non-IID data, that is, the participants may have  heterogeneous local data in terms of both size and distribution.
To address these problems, we exploit the meta-learning framework~\cite{vilalta2002perspective} that was first proposed to address the multi-task learning problem~\cite{zhang2017survey}, which can quickly adapt to a new task with only a small amount of training data.
The rationale is that as we may consider making a recommendation for each client as a task, we can estimate an inactive user's preferences even when only a small number of items have been consumed. 
In particular, we build PrivRec upon the gradient-based meta-learning framework Model-Agnostic Meta-Learning (MAML)~\cite{finn2017model}, which works purely by gradient-based optimization without requiring
additional parameters or model modification. 
In what follows, we give some preliminaries of MAML, then describe how PrivRec addresses data heterogeneity with the notion of meta-learning. 

\subsubsection{Model-Agnostic Meta-Learning (MAML)}
\label{st:MAML}

Given a model parameterized by $\theta$, and some related tasks (supervised or unsupervised), MAML runs several training rounds until the model converges or some condition is met.

In each round, the current global model parameter is denoted  as $\theta_0$, and the following steps are sequentially performed to further optimize  $\theta_0$.

\textbf{(1) Task Sampling:} 
A mini-batch of $M$ tasks (i.e., clients in our case) $\mathcal{T}_i$ are uniformly sampled. 

\textbf{(2) Local Update:}
Within each task $\tau \in \mathcal{T}_i$, the model parameter denoted as $\theta_\tau$ is initialized to be $\theta_0$:
\begin{equation}
	\theta_\tau = \theta_0.
\end{equation}
Also, its own data is split into support set $\mathcal{D}^s_\tau$ and query set $\mathcal{D}^q_\tau$.
Then $\theta_\tau$ is updated by gradient  $\nabla_{\theta_\tau} \mathcal{L}(\theta_\tau, \mathcal{D}^s_\tau)$ where $\mathcal{L}(\theta_\tau, \mathcal{D}^s_\tau)$ denotes the training loss on support set w.r.t. $\theta_\tau$.  Finally, we obtain the locally updated parameter $\theta_\tau$:
\begin{equation}
	\label{eq:localupdate}
	\theta_\tau \leftarrow \theta_\tau - \alpha_1 \nabla_{\theta_\tau} \mathcal{L}(\theta_\tau, \mathcal{D}^s_\tau),
\end{equation}
where $\alpha_1$ is the learning rate. Note that there could be more than one gradient descent step, and we show the case with only one step.

\textbf{(3) Meta Update:} After local updates, the initial parameter $\theta_0$ is globally updated based on the gradients of  $\mathcal{L}(\theta_\tau, \mathcal{D}^q_\tau)$, i.e.,  the meta-loss for task $\tau$ using the local updated parameter $\theta_\tau$ on query set $\mathcal{D}^q_\tau$:
\begin{equation}
	\label{eq:updatemaml}
	\theta_0 \leftarrow \theta_0 -\alpha_2 \nabla_{\theta_0 } \sum_{\tau \in \mathcal{T}_i} \mathcal{L}_\mathrm{DSSM}(\theta_\tau, \mathcal{D}^q_\tau),
\end{equation}
where $\alpha_2$ is the learning rate and the summation is over all sampled tasks $\tau \in \mathcal{T}_i$.

When training is completed, we obtain the optimal model parameter $\theta^\ast$.
Given an unseen task, starting from the optimal  parameter $\theta^\ast$, the model can quickly adapt to it by taking a small number of local gradient descent steps  using the support set (i.e., Eq.(\ref{eq:localupdate})).

\subsubsection{Faster-adapting PrivRec for Federated Learning}
Naturally, it is plausible to consider a client as a task in the context of FL, and implement the techniques of meta-learning.
Recently, \cite{chen2018federated} proposed a federated framework that integrates the aforementioned MAML for recommendation, in which a parameterized meta-algorithm is used to train the recommendation models, and parameters within both the local models and the meta-algorithm need to be optimized.
However, it is subject to inevitable deficiencies due to its straightforward application of the vanilla MAML.
First, the MAML meta-gradient update involves a gradient-of-gradient (also known as second-order gradient) calculation, which can be computationally expensive~\cite{finn2017model}. This also creates potentially infeasible memory requirements, especially for resource-constrained personal devices~\cite{wang2020next}. 
In~\cite{rajeswaran2019meta}, it is theoretically and empirically shown  that the compute cost consumed by First-order MAML (FOMAML) increases quite stably as the gradient steps increases, since FOMAML requires only gradient computations. Compared to FOMAML, the compute cost of MAML grows at a faster rate because the back-propagating through gradient descent of MAML requires Hessian-vector products at each iteration, which are more expensive. Similar patterns could be observed for computation time needed.
Second, it requires the split of local data for each client into support and query sets for the two-stage update, which may be impossible for inactive users with very few historical records. Moreover, in recommendation tasks, the majority of users are inactive because the number of user interactions commonly follows the long-tail distribution~\cite{yin2019social}.

To address the problems, we develop a faster-adapting mechanism to PrivRec based on the REPTILE algorithm~\cite{nichol2018first}. As an approximation of MAML, REPTILE executes stochastic gradient descent for a fixed number of iterations on a given task, and then gradually moves the initialization weights in the direction of the weights obtained from the tasks. By ignoring the second-order gradients, REPTILE avoids the heavy computation and needs to split the local data while still achieving promising results in the meta-learning task~\cite{nichol2018first}. Motivated by this, we reformulate the meta update in Eq.(\ref{eq:updatemaml}) to be:
\begin{equation}
	\label{eq:updatereptile}
	\theta_0 \leftarrow \theta_0 + \alpha_2 \frac{1}{M} \sum_{\tau=1}^M (\theta_\tau - \theta_0),
\end{equation}
where $\alpha_2$ is the learning rate, $M$ is the number of sampled tasks and $\theta_\tau$ denotes the locally updated parameters in task $\tau$. This is different from Eq.(\ref{eq:updatemaml}) since it does not need to take any derivatives from a different dataset.
This is also different from the ordinary SGD as we allow for multiple local update iterations to obtain $\theta_\tau$, which is the key to better on-device personalization in practice \cite{jiang2019improving}.


With these ingredients, we formally list the details of training PrivRec in Algorithm 1, where steps executed on the server side (i.e., task sampling and meta update) and client side (i.e., local update) are respectively summarized in lines 2-8 and 10-18.

\begin{algorithm}[t!]
	
	\caption{Procedures for Training PrivRec} \BlankLine
	\KwIn{The number of global rounds $\ensuremath{E}_1$ and local epochs $\ensuremath{E}_2$, the number of clients sampled each time $M$, initialized model parameter set $\theta_0 = \{ \mathbf{U}, \mathbf{V}, \mathbf{W}, \mathbf{b} \}$, constant $\mu$ and learning rates $\alpha_1, \alpha_2$;} 
	\textbf{\textsc{[Server Execution]}:} \\
	
	\For{round $i \leftarrow 1, 2, \dots, \ensuremath{E}_1$}
	{
		Randomly sample $M$ clients  to form a task set $\mathcal{T}_i$;\\
		\For{client $\tau \in \mathcal{T}_i$ }
		{
			$\theta_{\tau} \leftarrow \textsc{ClientUpdate}(\tau, \theta_0)$;\\
		}
		$\theta_0 \leftarrow \theta_0 + \alpha_2 \frac{1}{M} \sum_{\tau=1}^M \Delta_\tau$;\\	
	}
	\BlankLine	
	\textbf{\textsc{[Client Execution, i.e.,}} $\textsc{ClientUpdate}(\tau, \theta_0)$\textbf{\textsc{]}:}\\
	$\mathcal{B}_\tau \leftarrow$ (divide the local data into mini-batches) ;\\
	$\theta_\tau \leftarrow \theta_0$;\\
	\For{$j \leftarrow 1, 2, \dots, \ensuremath{E}_2$}
	{
		\For{batch $\mathcal{B}'_\tau \in \mathcal{B}_\tau$}
		{
			$\theta_\tau\leftarrow\theta_\tau - \alpha_1 \nabla_{\theta_\tau} \mathcal{L}'(\theta_\tau,\mathcal{B}_\tau')$;\\
		}

	}
	$\Delta_\tau = \theta_\tau - \theta_0$;\\

	return $\Delta_\tau$ to the server;\\
	\label{alg:fedavg}
\end{algorithm}

\subsection{Differentially Private-PrivRec (DP-PrivRec)}
So far, by running PrivRec in the FL setting, we ensure that user data is retained in the local device, facilitating privacy protection regarding personal data.
However, risk still exists if there are potential participating clients who are able to infer whether a given user is present during training solely based on the model parameters or gradients they receive~\cite{melis2019exploiting,orekondy2018gradient}, thus leaking user identities and even sensitive attributes. 

To defend PrivRec against such attacks, we bridge the model with the notion of differential privacy  (DP)~\cite{dwork2006calibrating,abadi2016deep}, which is the state-of-the-art framework for quantifying and limiting information disclosure about individuals due to its strong privacy guarantee.
We adopt DP because of its several unique advantages. First, it provides strong theoretical privacy guarantees without any information about an individual and is not affected by what the attackers know about the dataset. The benefit is that the involved parties can access datasets without ad-hoc restrictions, maintaining the advantages of the  FL training  framework.  Second, a very nice feature of DP is that it can be composed. Specifically, in modern machine learning training process, many times of access to training data are required, which would naturally result in more data exposure. Theoretically, the composition feature of DP enables it to achieve minimized privacy loss in various settings when accessing dataset many times. We can use a wide range of privacy composition techniques~\cite{jayaraman2019evaluating} to keep track of and limit the privacy loss when training our proposed deep learning-based recommender system.

A conventional DP mechanism usually introduces a level of uncertainty into the released data, such that the contribution of any data point will not lead to obvious changes in the data. However, in the FL setting, instead of only protecting a single training data point, we need to protect each user's entire dataset from attacks. Hence, we extend PrivRec with our proposed user-level DP mechanism, which is named DP-PrivRec and is introduced in this section.

\subsubsection{Differential Privacy (DP)}
We hereby present some necessary preliminaries regarding DP. The \textit{adjacent datasets} is the core concept in DP. Two datasets are considered adjacent when they are identical except for one record. With respect to a \textit{record}, most prior work such as ~\cite{abadi2016deep} and~\cite{phan2017adaptive} tend to derive their task-specific definitions, such as a single training sample, a mini-batch of training samples, or all the data from a single user. 
As we aim to enforce the user-level privacy, the adjacent datasets in our paper are defined below.
\begin{definition}[\textbf{Adjacent Datasets}]
	Let $d$ and $d'$ be two datasets where each entry is associated with a user. $d$ and $d'$ are \textit{adjacent} if we can obtain $d'$ by replacing all data points associated with only one user in $d$ with the examples of a different user from $d'$.
\end{definition}
Different from the definition in \cite{mcmahan2017learning} that would lead to a variable size of mini-batch  of clients at each training round, this definition instead yields a fixed mini-batch size. This makes it easier to analytically compute the privacy loss by using the composition rule~\cite{mcmahan2018general}, and further accelerate computation.

Recall that our goal is to ensure that the presence or absence of any specific user's data in the training set should have slight impact on the parameters of the learned model, which means that it is impossible for an adversary to infer whether any specific user's data has been used in the training data (i.e., whether it is $d$ or $d'$) by examining the trained model. This goal can be formally expressed as $(\eps,\delta)$-Differential Privacy below.
\begin{definition}[\textbf{$(\eps,\delta)$-Differential Privacy}]
	\label{def:dp}
	A randomizing mechanism $\M\colon \Domain\rightarrow\Range$ with domain $\Domain$ and range $\mathcal{R}$ 
	satisfies $(\eps,\delta)$-differential privacy if for any two adjacent inputs $\D,\D'\in \Domain$ and for any subset of outputs $O\subseteq\Range$, it holds that
	$\Pr[\M(\D)\in O]\leq e^{\eps}\Pr[\M(\D')\in O]+\delta$ where $Pr[\cdot]$ denotes the probability.
\end{definition}

There are a wide range of mechanisms to quantify data privacy such as Laplace Mechanism and Gaussian Mechanism. The privacy loss random variable after composing multiple Gaussian Mechanisms also follows a Gaussian distribution, and this nice feature makes composition analysis in the training of deep models more friendly. As a result, given the same privacy budget, we can achieve tighter privacy guarantee by composing multiple Gaussian Mechanisms in deep learning~\cite{jayaraman2019evaluating}.

To achieve  $(\eps,\delta)$-DP using Gaussian Mechanism, a common method is to add noise drawn from the Gaussian distribution to the output of mechanism $\M$.
Note that in DP-PrivRec, the outputs of $\M$ from the clients are gradients, so
we leverage noisy stochastic gradient descent (NoisySGD), which is a popular option when deploying DP-enhanced DNNs~\cite{abadi2016deep,song2013stochastic,mcmahan2017learning}.
According to NoisySGD, within each iteration of model training, a gradient w.r.t. the model loss function on a randomly subsampled dataset is obtained and then the gradient is clipped (or bounded) in $\ell_2$ norm.
Clipping the gradients effectively bounds the sensitivity of the system w.r.t. the addition or removal of an arbitrary individual from the training set. After gradient clipping, the Gaussian noise perturbation on the gradients ensures $(\eps,\delta)$-DP for this iteration.

\subsubsection{Building DP-PrivRec}
McMahan \textit{et al.}~\cite{mcmahan2017learning} added a differential privacy mechanism to the FL optimization algorithm to achieve user-level privacy protection, and used moments accountant~\cite{abadi2016deep} to verify that it satisfies user-level differential privacy.
Motivated by this, we propose our user-level differentially private recommender DP-PrivRec.
Specifically, at the $i$-th round of training DP-PrivRec, we implement the following steps.

\textbf{(1) Subsampling Clients:}
Given $N$ clients, we uniformly sample a batch of $M$ clients $\mathcal{T}_i$ with probability $q$, where the ratio $q=M/N$ is defined as the sampling parameter and is important for measuring privacy loss.
It is also worth noting that subsampling can amplify the privacy guarantee~\cite{kasiviswanathan2011can} since it decreases the chances of leaking information about a particular individual, and makes it impossible to infer the information about this individual when she/he is not included in $\mathcal{T}_i$.

\textbf{(2) Clipping Gradient:}
On client $\tau$, we compute the gradients w.r.t. the loss function on mini-batches of local datasets. After each computation, we obtain a new gradient, denoted by $\Delta_\tau$, whose $\ell_2$ norm is bounded by a predefined threshold $S$:
\begin{equation}
	\Delta_\tau \leftarrow \Delta_\tau  \cdot \min\{1, \frac{S}{||\Delta_\tau||}\}.
\end{equation}
After gradient clipping, the gradient is sent back to the server.

\textbf{(3) Estimating Gradients:}
On the server side, these clipped gradients are aggregated on the central server using a query function, which is in accordance with Eq.(\ref{eq:updatereptile}): $f(\mathcal{T}_i) \coloneqq \frac{\sum_{\tau \in \mathcal{T}_i} \Delta_\tau}{M}$. \\

\textit{Theorem 1.} The sensitivity of $f(\mathcal{T}_i)$ is upper-bounded by $\frac{2S}{M}$. \\


\textit{Proof of  Theorem 1.} Given any two adjacent batch $\mathcal{T}_i$ and $\mathcal{T}_{i'}$, and assuming, by definition, that they are only different in user $u$ and user $u'$, we have:
\begin{align*}
	\label{eq:boundpf}
	\| f(\mathcal{T}) - f(\mathcal{T}_{i'})\| &= \|\frac{\sum_{\tau \in \mathcal{T}_i} \Delta_\tau}{M} - \frac{\sum_{\tau \in \mathcal{T}_{i'}} \Delta_\tau}{M} \| \\
	&= \| \frac{\Delta_u - \Delta_{u'}}{M} \| \\
	&\le \frac{2S}{M} 
\end{align*}
$\hfill \Box$

\textbf{(4) Adding Noise:}
Finally, we draw noise from the Gaussian distribution $N(0, \sigma^2)$, which is used to perturb the aggregated gradient, where $\sigma$ is proportional to the sensitivity of the query function, i.e., $\sigma = z \cdot \frac{2S}{M}$.

\subsubsection{A Privacy Guarantee}
\label{st:guarantee}
In this section, we will discuss a privacy guarantee of such mechanism.
In~\cite{abadi2016deep} it is discussed that privacy spent for multiple access to the sensitive data can be measured by the moments accountant technique.
Next, we demonstrate that our model is also applicable for moments accountant. \\

\textit{Proof that our model is applicable for moments accountant.}
(1) At each round, each batch with size $M$ is randomly sampled equally by the ratio $q$, which meets the condition of privacy amplification.
(2) The gradient from each client is clipped to $S$. 
(3) We have proved that the query for calculating new $\Delta$, namely $f(\mathcal{T})$, is upper-bounded by $\frac{2S}{M}$.
According to~\cite{mcmahan2018general}, if we apply the Gaussian mechanisms $(\frac{2S}{M}, z)$ in a training round, then the resulting Gaussian mechanism is with privacy tuple $(\frac{2S}{M}, z)$. 
(4) The moments are upper-bounded by that of the sampled Gaussian mechanism with sensitivity 1, noise scale $z$ and sampling probability $q$. 
(5) $z$ is used to generate the noise irrespective of the private data.
Hence, we can safely apply the composability property of moments accountant. $\hfill \Box$ \\

These Gaussian mechanisms are applied to $N/M$ subsamples sequentially, whose privacy losses are recorded by the moment accountant data structure.
Abadi \textit{et.al.}~\cite{abadi2016deep} apply composability and tail bound properties of these moment accountants to obtain $\epsilon$ given budget $\delta$, achieving $(\epsilon, \delta)$-DP.

In our work, we employ a numerically  stable  analytical Moment Accountant implementation named autodp~\footnote{https://github.com/yuxiangw/autodp}, which is based on Renyi Differential Privacy (RDP)~\cite{wang2019subsampled}, to calculate the privacy spent.

\begin{algorithm}[h]
	\SetAlgoLined
	\caption{Procedures for Training DP-PrivRec} \BlankLine
	\KwIn{The size of each batch for global update $M$, the number of global and local training epochs $\ensuremath{E}_1$ and $\ensuremath{E}_2$, the initialized model parameter set $\theta_0 = \{ \mathbf{U}, \mathbf{V}, \mathbf{W}, \mathbf{b} \}$  where $\mathbf{V}$ is from the first stage, noise scale $z$ for Gaussian Mechanism, clipping bound $S$, learning rates $\alpha_1, \alpha_2$;} 
	\textbf{\textsc{[Server Execution]}:} \\
	\For{each round $i = 1, 2, \dots, {E}_1$}
	{
		Select a batch of  $M$ clients $\mathcal{T}_i$ from a total of $N$ clients, with the ratio $q=M/N$;\\
		\For{each client $\tau \in \mathcal{T}_i$ \textbf{in parallel}}
		{
			$\Delta_{\tau} = \textsc{ClientUpdate}(\theta_0, S)$;\\
		}
		
		$\Delta = \frac{\sum_{\tau \in \mathcal{T}_i} \Delta_\tau}{M}$;\\ $\sigma_{\Delta} = \frac{2S}{M} \cdot z $;\\
		$\theta_0 \leftarrow \theta_0 + \alpha_2 \Delta + \mathcal{N}(0, I\sigma_{\Delta}^2)$;\\
		
	}
	\BlankLine	
	\textbf{\textsc{[Client Execution, i.e.,} [\textsc{Client Update}($\theta_0, S$)]:}\\
	$\mathcal{B} =$ (split the local data into batches) ;\\
	$\theta_\tau  = \theta_0$;\\
	\For{each local epoch $j = 1, 2, \dots, \ensuremath{E}_2$}
	{
		\For{batch each $\mathcal{B}' \in \mathcal{B}$}
		{
			\If{two-stage training}
			{
			$\theta_\tau \leftarrow \theta_\tau - \alpha_1 \nabla_{\theta_\tau} \mathcal{L}(\theta_\tau,\mathcal{B}')$;\\
			}
			\If{one-stage training}
			{
				$\theta_\tau \leftarrow \theta_\tau - \alpha_1 \nabla_{\theta_\tau} \mathcal{L}_\mathrm{DSSM}(\theta_\tau,\mathcal{B}')$;\\
			}
			
		}
		$\Delta_\tau = \theta_\tau  - \theta^0$;\\
		$\Delta_\tau  \leftarrow \Delta_\tau \cdot \min\{1, \frac{S}{||\Delta_\tau||}\}$;\\
		
	}

	return $\Delta_\tau$ to the server;\\
	\label{alg:fedavg}
\end{algorithm}

\subsection{Two-stage FL Training}
Although adding noises during FL model training could further boost privacy protection, it is natural and inevitable to see degraded performance.
To address this problem, we propose a two-stage FL training approach.
In the first stage, we only focus on learning of item representations using a self-supervised learning method (SSL) and do not impose the differential privacy mechanism during training.
This is based on the observation that we could relax the level of privacy protection when user information is exclusive, which enhances the item representation learning.
In the second stage, we follow the training procedures of DP-PrivRec and optimize the model parameters by a downstream task, namely, modelling the user-item interactions.
We employ the well-trained item embeddings from the last stage as initialized values, and use the labeled user-item interactions to learn user embeddings and fine-tune item embeddings.

Next, we discuss how to utilize SSL to effectively learn item representations in the first stage.

\subsubsection{Self-supervised Learning}
In the first stage, we aim to learn item representations without involving user information. However, this would be challenging when labels (i.e., user-item interaction in this case) are absent.
To this end, we turn to the emerging self-supervised learning (SSL) paradigm to tackle this dilemma, which has shown success in CV~\cite{kolesnikov2019revisiting,doersch2015unsupervised,caron2018deep} and NLP~\cite{le2014distributed,devlin2018bert}.
When labeled data is unavailable, SSL designs a domain specific pretext task, which assigns different labels for data instances and therefore learns from unlabeled data itself. 
When there is also limited labeled data available, SSL can be used a pre-training process, after which the labeled data can be used to fine-tune the deep model in a downstream task.

\subsubsection{Initializing Item Representations with SSL}

Since we do not rely on user information in this stage, the user-item interaction as supervision signals cannot be utilized. Instead, we design a SSL pretext task to learn their representations based on inter-item relations. In our case, we consider the temporally ordered item sequences interacted by user (e.g., clicked or liked).  In some studies, they are also called sessions~\cite{hidasi2015session,hidasi2016parallel}. We utilize such input due to two main reasons. First, the sequences are naturally formed based on the temporal order  without including user's identifications, which avoids the dependencies of user representation learning in this stage. Second, the sequence itself as a commonly seen structure  can  provide rich and robust auxiliary self-supervision signals.

Given a sequence of orderly interacted items, we enable the pretext task to capture the item transitions among items. Through optimizing the learning objective of the pretext task, we can obtain the item representations.
Although the problem of session-based recommendation has been extensively studied~\cite{wang2019survey}, many of them explicitly model user preferences (e.g., long- and short-term preferences) from sessions~\cite{guo2019streaming,wang2019collaborative}, exploit cross-session features~\cite{xia2020self,qiu2020exploiting} or design a global set to aggregate frequent items~\cite{wang2020global}, in order to improve the session-based recommendation performance. These approaches are not suitable for the unique environment of federated learning.

Formally, our SSL pretext task employs contrastive learning ~\cite{oord2018representation}, a popular framework for SSL. Its major idea is to derive multiple views from  the original input data and maximize the mutual information between encoded representations of these views.

In device $d$, given an activity sequence $X$  that contains $T$ consecutive visited items by a user within a certain time interval $X = [v_1, v_2,\ldots, v_T]$, we derive two different views of $X$ inspired by BERT~\cite{devlin2018bert} and its extension~\cite{kong2019mutual}:
\begin{description}
	\item[\textbf{Item Masked Sequence} ] We  mask the $i$-th item $v_i$ in the original sequence $X$ by replacing it with a masked item $\hat{v}_i$ to be  $\hat{X}_i = [v_1, \ldots, \hat{v}_i,\ldots, v_T]$, where $\hat{v}_i$ is a negative item and can be randomly drawn from the item sets. 
	\item[\textbf{Segment Masked Sequence} ] We also mask a segment $v_{i:j}$ spanning from  position $i$ to $j$  in the original sequence $X$ by replacing it with another segment $\bar{v}_{i:j}$:  $\bar{X}_{ij} = [v_1, \ldots,\bar{v}_{i:j},\ldots, v_T]$, where $\bar{v}_{i:j}$ is a negative segment.
\end{description}

With these derived views, we formulate the SSL pretext task learning objectives.

\textbf{Target Item Prediction}
The first one is based on an item masked sequence $\hat{X}_i$ and the $i$-th item to be masked $v_i$, and predicts the target positive item $v_i$ using
\begin{equation}
\label{eq:lim}
\begin{aligned}
\mathcal{L}_\mathrm{IM} &= \mathbb{E}_{p(\hat{X}_i, v_{i})}[f_{\boldsymbol{\Sigma}}(\hat{X}_i)^{\top} f_{\boldsymbol{\psi}}(v_{i}) \\
&-\log \sum_{\hat{v}_i \in \hat{V}} \exp (f_{\boldsymbol{\Sigma}}(\hat{X}_i)^{\top} f_{\boldsymbol{\psi}}(\hat{v}_i))],
\end{aligned}
\end{equation}
where $f_{\boldsymbol{\Sigma}}$ is an encoder parameterized by $\Sigma$ to learn sequence representations, $f_{\boldsymbol{\psi}}$ is an encoder parameterized by $\boldsymbol{\psi}$ to learn item representations, $\hat{V}$ is the item set from which negative samples are drawn.
Here we use a GRU encoder~\cite{hidasi2015session} as $f_{\boldsymbol{\Sigma}}$ to encode a session, and use the hidden state at the last time step to represent it. Also, we use a feed-forward network as $f_{\boldsymbol{\psi}}$ to learn item representations.

The rational behind this learning objective is to maximize the scores between positive pair $(\hat{X}_i, x_{i})$ and negative pairs $(\hat{X}_i, \hat{v}_i)$.

\textbf{Target Segment Prediction}
Similarly, the second learning objective $\mathcal{L}_\mathrm{SM}$ is based on a segment masked sequence $\bar{X}_{i:j}$ and a positive segment $v_{i:j}$, and aims to maximize the scores between the true and negative pairs, namely,  $(\bar{X}_{i:j}, v_{i:j})$ and $(\bar{X}_{i:j}, \bar{v}_{i:j})$:
\begin{equation}
\label{eq:lsm}
\begin{aligned}
\mathcal{L}_\mathrm{SM} &= \mathbb{E}_{p(\bar{X}_{i:j}, v_{i:j})}[f_{\boldsymbol{\Sigma}}(\bar{X}_{i:j})^{\top} f_{\boldsymbol{\Sigma}}(v_{i:j}) \\
&-\log \sum_{\bar{v}_{i:j} \in \bar{S}} \exp (f_{\boldsymbol{\Sigma}}(\bar{X}_{i:j})^{\top} f_{\boldsymbol{\Sigma}}(\bar{v}_{i:j}))],
\end{aligned}
\end{equation}
where $f_{\boldsymbol{\Sigma}}$ is the same sequence encoder with that in Eq~\ref{eq:lim},  $\bar{S}$ is the negative segment set and can be obtained by randomly sampled from other sessions of this user or other positions of this session.

Therefore, the overall SSL learning objective is the combination of $\mathcal{L}_\mathrm{IM}$ and $\mathcal{L}_\mathrm{SM}$, and can be formulated as:
\begin{equation}
\mathcal{L}_\mathrm{SSL} = \lambda_\mathrm{IM}\mathcal{L}_\mathrm{IM} + \lambda_\mathrm{SM}\mathcal{L}_\mathrm{SM},
\end{equation}
where $\lambda_\mathrm{IM}$ and $\lambda_\mathrm{SM}$ are two hyper-parameters to balance their contributions. 

Following the training procedures of  FL, we learn the item representations based on $\mathcal{L}_\mathrm{SSL}$ without adding any DP noises.

\subsubsection{Fine Tuning Item Representations and Learning User Representations}
In the second stage, we use the item  representations in the first stage as initialization and optimize the model parameters following DP-PrivRec.
Note that the objective function now becomes  a joint learning $\mathcal{L}$:
\begin{equation}
\mathcal{L} = \lambda_\mathrm{DSSM} \mathcal{L}_\mathrm{DSSM} + \mathcal{L}_\mathrm{SSL},
\end{equation}
where $\lambda_\mathrm{DSSM}$ is the hyper-parameter to control the contribution of $\mathcal{L}_\mathrm{DSSM}$ in Eq.(\ref{eq:dssm}).

\section{Evaluation Setup}
\label{se:exp}
We introduce our evaluation settings in this section.

\begin{table*}[h]
		\center
	\caption{Comparison of Our Proposed Models and Relevant Baselines on Movielens.}
\begin{tabular}{|l|l|l|l|l|l|l|l|l|}
	\hline
	\multicolumn{1}{|c|}{Metric}           & \multicolumn{4}{c|}{Hits@k}   & \multicolumn{4}{c|}{nDCG@k}   \\ \hline
	\multicolumn{1}{|c|}{k}                & 5     & 10    & 20    & 30    & 5     & 10    & 20    & 30    \\ \hline
	FOFL-SELF                              & 0.118 & 0.152 & 0.195 & 0.241 & 0.101 & 0.130 & 0.169 & 0.207 \\ \hline
	NN-SELF                                & 0.109 & 0.148 & 0.190 & 0.235 & 0.098 & 0.124 & 0.162 & 0.201 \\ \hline
	FCF                                    & 0.092 & 0.140 & 0.186 & 0.228 & 0.088 & 0.118 & 0.155 & 0.195 \\ \hline
	FedGNN                                 & 0.120 & 0.155 & 0.201 & 0.247 & 0.109 & 0.137 & 0.174 & 0.218 \\ \hline
	NCF                                    & 0.158 & 0.206 & 0.294 & 0.347 & 0.152 & 0.186 & 0.250 & 0.307 \\ \hline
	\textbf{PrivRec}     & 0.131 & 0.168 & 0.217 & 0.269 & 0.114 & 0.141 & 0.189 & 0.233 \\ \hline
	\textbf{PrivRec-CEN} & 0.136 & 0.187 & 0.235 & 0.287 & 0.125 & 0.152 & 0.206 & 0.258 \\ \hline
\end{tabular}
\label{tb:overall1}
\end{table*}

\begin{table*}[h]
	\center
	\caption{Comparison of Our Proposed Models and Relevant Baselines on Frappe.}
\begin{tabular}{|l|l|l|l|l|l|l|l|l|}
	\hline
	\multicolumn{1}{|c|}{Metric} & \multicolumn{4}{c|}{Hits@k}   & \multicolumn{4}{c|}{nDCG@k}   \\ \hline
	\multicolumn{1}{|c|}{k}      & 5     & 10    & 20    & 30    & 5     & 10    & 20    & 30    \\ \hline
	FOFL-SELF                    & 0.536 & 0.618 & 0.655 & 0.682 & 0.479 & 0.554 & 0.600 & 0.621 \\ \hline
	NN-SELF                      & 0.531 & 0.613 & 0.651 & 0.677 & 0.472 & 0.550 & 0.595 & 0.616 \\ \hline
	FCF                          & 0.527 & 0.506 & 0.646 & 0.671 & 0.469 & 0.544 & 0.591 & 0.612 \\ \hline
	FedGNN                       & 0.541 & 0.625 & 0.672 & 0.688 & 0.486 & 0.563 & 0.617 & 0.631 \\ \hline
	NCF                          & 0.598 & 0.658 & 0.712 & 0.770 & 0.547 & 0.612 & 0.663 & 0.698 \\ \hline
	\textbf{PrivRec}             & 0.549 & 0.635 & 0.679 & 0.697 & 0.495 & 0.570 & 0.621 & 0.636 \\ \hline
	\textbf{PrivRec-CEN}         & 0.564 & 0.641 & 0.690 & 0.726 & 0.501 & 0.574 & 0.632 & 0.651 \\ \hline
\end{tabular}
\label{tb:overall2}
\end{table*}

\subsection{Experimental Environment}
Our experiments follow the FL environment simulation widely used in FL research~\cite{ma2020safeguarding,jalalirad2019simple}.
Specifically, we build our models using the popular PySyft~\cite{ryffel2018generic} FL framework, where each user is modeled as a virtual worker object and behaves exactly like normal remote edge devices.
A virtual server object is also created for conducting the global model update and controlling the training process. By following the standard simulation setting, we focus on the core logic of our problem in a real-life production scenario.

\subsection{Datasets}
\label{st:dataset}
In this section, we introduce the two datasets Frappe and Movielens-1M, which have ample user and item features for representation learning. We introduce their properties as follows:

\textbf{Frappe}\footnote{http://baltrunas.info/research-menu/frappe}:
Frappe is a context-aware mobile application discovery tool. We adopt the extended version of the dataset used in~\cite{he2017neuralCF}. The dataset contains eight features, which are ``count'', ``daytime'', ``weekday'', ``is weekend'', ``homework'', ``cost'', ``weather'', ``country'' and ``city''.
We consider ``count'' and ``cost'' as item-related features while the rest as user features. We convert the ratings greater than 3 stars as positive ratings (i.e., 1) and the others as negative ratings (i.e., 0).  This dataset contains 957 users, 4,082 items  and 288,609 interactions.

\textbf{Movielens}\footnote{https://grouplens.org/datasets/movielens/}: Movielens provides datasets containing the movie ratings by users. We select Movielens-1M in this experiment, which contains user demographic information: age, gender and occupation, and 18 movie genres. Similarly, we denote every feature (including each genre) as a one-hot feature vector. Similarly,  we consider the genres as item related features and user demographic information as user features. We binarize the user ratings into implicit feedbacks via the same strategy as used for Frappe. This dataset contains 6,040 users, 3,706 items and 1,000,209 interactions.

\begin{figure}[!t]
	\centering
	\includegraphics[scale = 0.6]{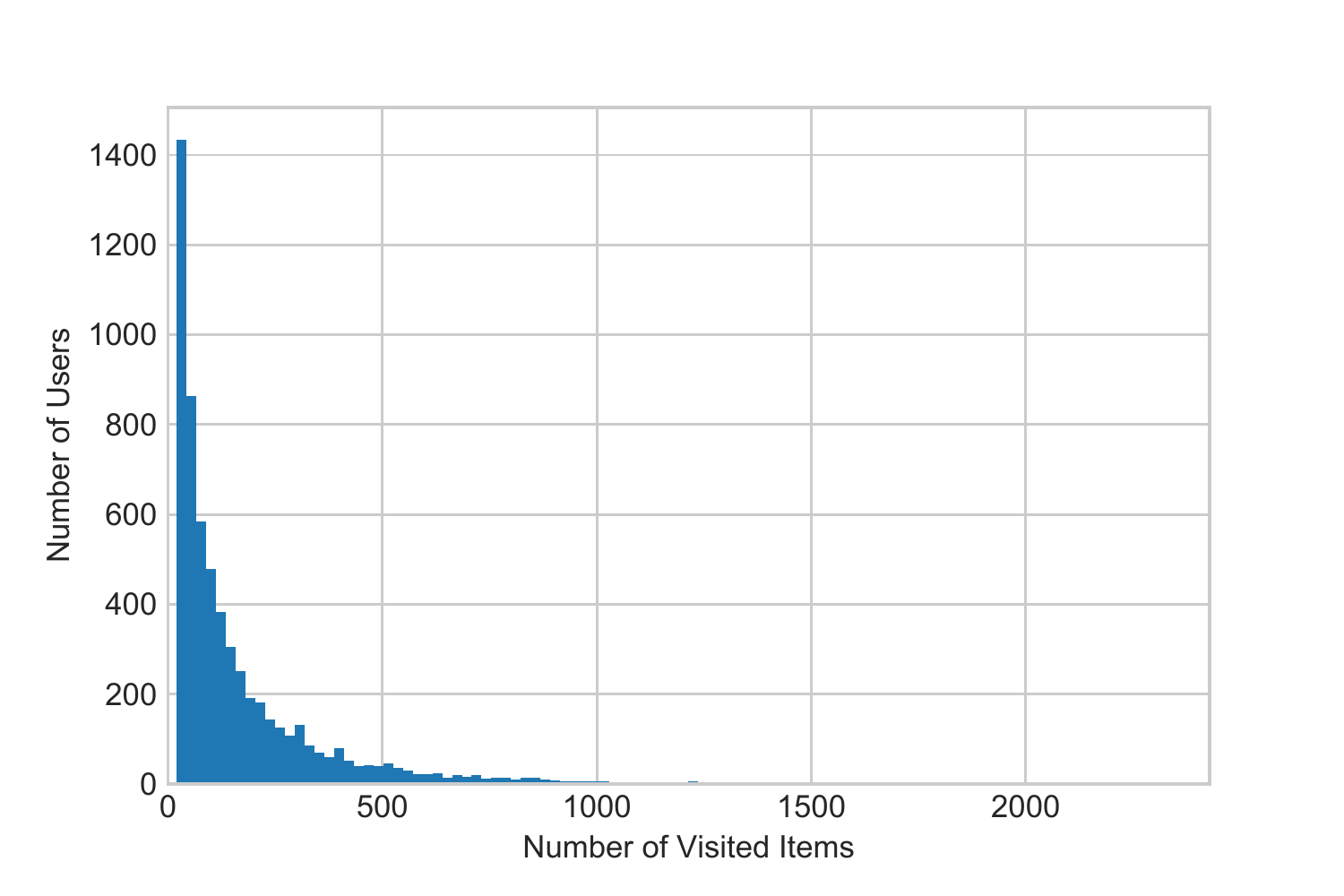}
	\caption{\textbf{The long-tail distribution of users with different number of visited items.}}
	\label{fig:exp:hist}
\end{figure}

To further demonstrate the need for fast personalization with few data points, we show how the number of user activities vary significantly in the Movielens dataset descriptions in Fig.~\ref{fig:exp:hist}. We can observe the long-tail nature of such distribution from the figure, and most users only visited a small number of items. The similar pattern can also be observed in the Frappe dataset.

%

\subsection{Evaluation Protocols}
We randomly select  80\% of the users as \textit{training users} and the rest 20\% as \textit{testing users}. 
Similar to the settings in prior work~\cite{chen2018federated,jalalirad2019simple}, we use the following method to split our datasets.
In the training phase, we use all the data from training users to train the model. After obtaining a fully pretrained model after the training phase, we move to the testing phase. Specifically, for each testing user/client $\tau$, we evenly split her/his data into the training set $\mathcal{D}^\tau_{train}$ and testing set $\mathcal{D}^\tau_{test}$.
For each user $\tau$, we fine-tune the pretrained model on the client device using  $\mathcal{D}^\tau_{train}$ to facilitate on-device personalization, and then evaluate the recommendation performance on $\mathcal{D}^\tau_{test}$ from all testing users.

To measure the recommendation accuracy, we use two evaluation metrics commonly used in recommender system research~\cite{cremonesi2010performance,wang2018neural,wang2020next}, namely hits ratio at rank $k$ (\textit{Hits@k}) and normalized discounted cumulative gain at rank $k$ (\textit{nDCG@k}). Specifically, for each positive user-item interaction in the test set $\mathcal{D}^\tau_{test}$, we proceed as follows:

(1) We compute ranking scores for the positive item as well as the negative items that the user has never interacted with.

(2) We form a top-$k$ recommendation list by picking $k$ items with the highest ranking scores. If the ground-truth item appears in the top-$k$ recommendation list, we have a \emph{hit}. Otherwise, we have a \emph{miss}.

Then, \textit{Hits@k} for each $\mathcal{D}^\tau_{test}$ is defined as:
\begin{equation}
	\label{eq:hits}
	\centering
	Hits@k=\frac{\#hit@k}{|\mathcal{D}^\tau_{test}|},
\end{equation}
where $\#hit@k$ denotes the number of \emph{hits} in all testing cases from user $\tau$. A high $Hits@k$ value is expected for a good recommender model. Meanwhile, for $\mathcal{D}^\tau_{test}$, \textit{nDCG@k} is defined as:
\begin{equation}
	\label{eq:ndcg}
	nDCG @ k= \sum_{i=1}^{k} \frac{2^{r_{i}}-1}{\log _{2}(i+1)},
\end{equation}
where $r_i$ is the graded relevance of item at position $i$. We use the simple binary relevance in the experiments, meaning that $r_i = 1$ if the ground-truth item $v$ is in the \emph{hits} set and $r_i = 0$ otherwise. After evaluating on each individual user, the \textit{Hits@k} and \textit{nDCG@k} scores are averaged over all testing users as the overall results.

\section{Experimental Results and Discussions}\label{st:result}
Following the settings in Section \ref{se:exp}, we conduct experiments to evaluate our proposed models. Specifically, we aim to verify our major claims made in this paper by answering the following research questions (RQs):

\noindent\textbf{RQ1.} Without DP, how does PrivRec perform compared with similar state-of-the-art models?

\noindent\textbf{RQ2.} What are the effects of different FL hyperparameter settings on the performance of PrivRec?

\noindent\textbf{RQ3.} How is the performance of PrivRec different from DP-PrivRec, which has added noise to enhance privacy?  Furthermore, does the two-stage approach outperform the one-stage approach in recommendation accuracy?

\noindent\textbf{RQ4.} How do the DP parameters affect the privacy guarantee? How does the recommendation accuracy of DP-PrivRec change as we vary the privacy budget?

\noindent\textbf{RQ5.} Can meta-learning help the global recommendation model effectively personalize to inactive users who  hold only few records? Is the first-order meta-learning method more computationally efficient than the second-order meta-learning method?

\noindent\textbf{RQ6.} Can DP-PrivRec effectively protect from membership inference attacks?

\begin{figure*}[t]
	\begin{tabular}{ccc}
		\includegraphics[scale=0.42]{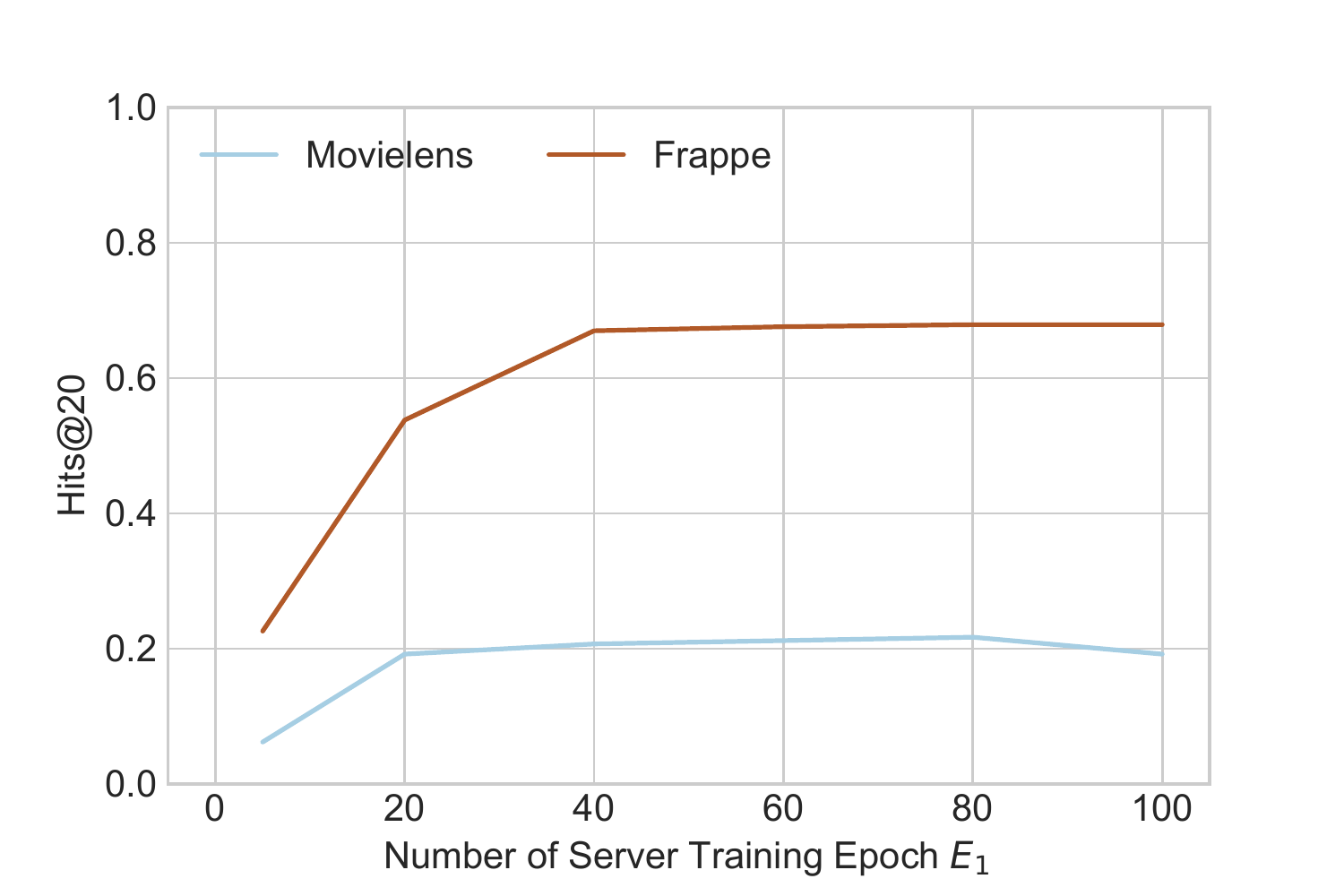} &  \includegraphics[scale=0.42]{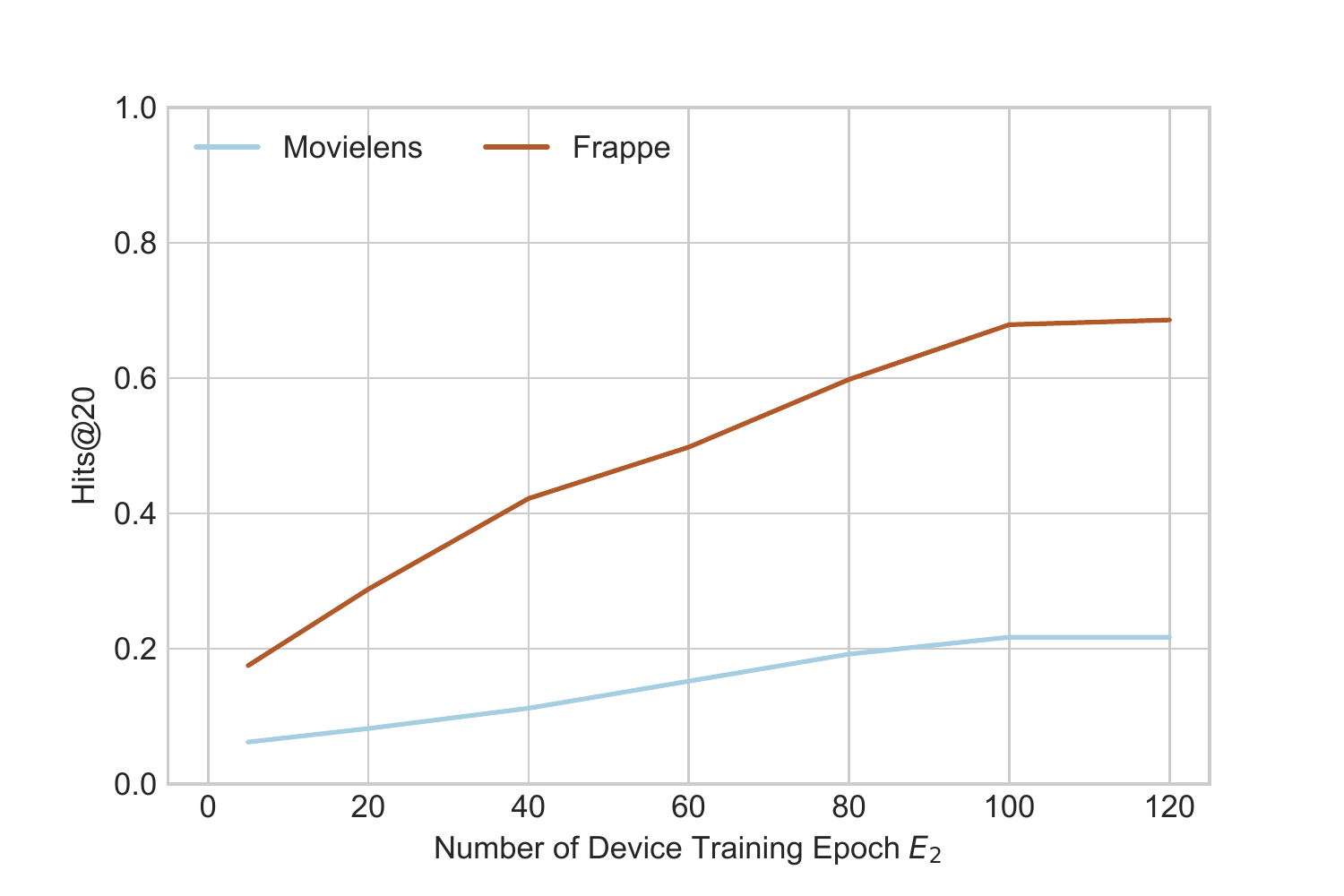} & \includegraphics[scale=0.42]{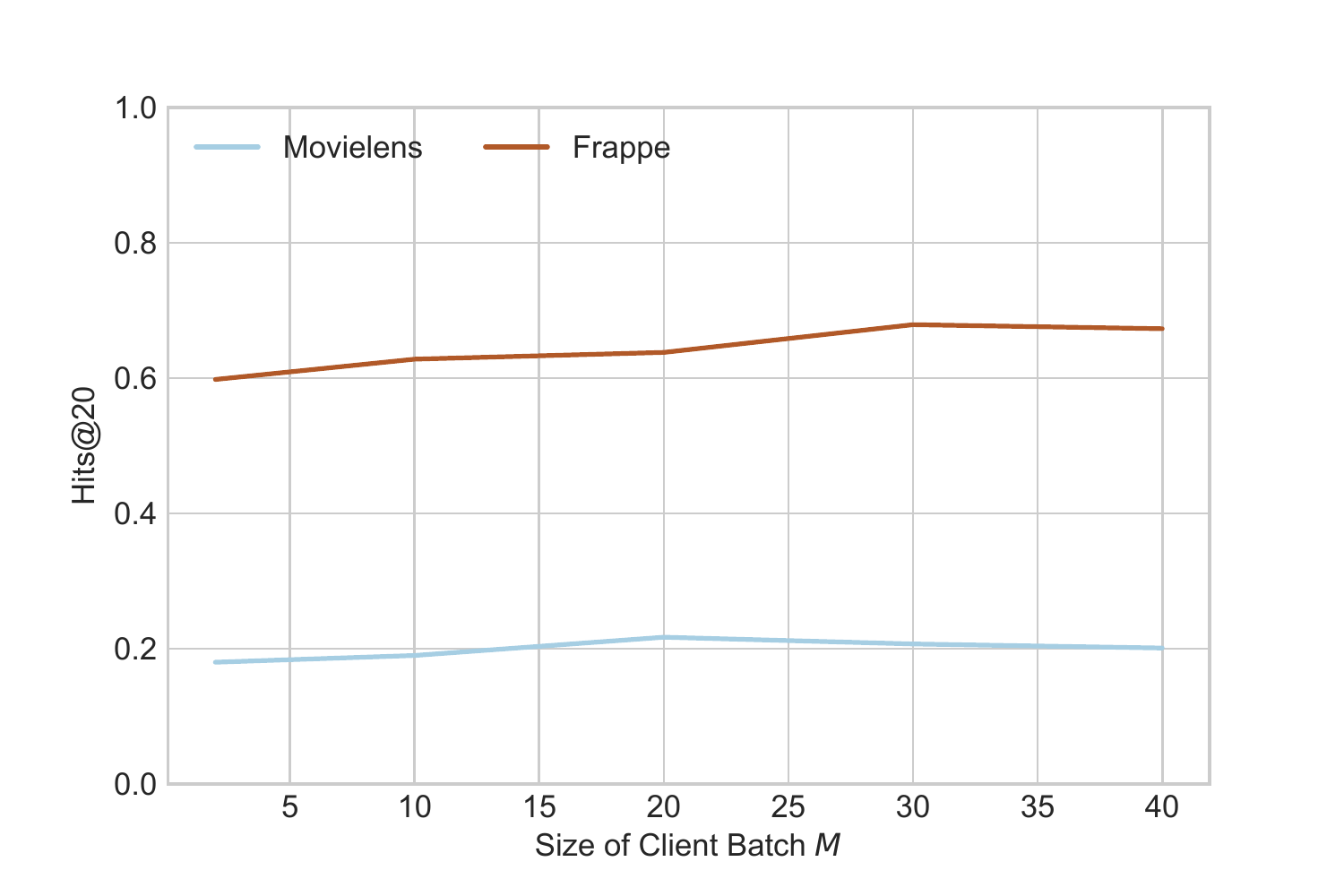} \\
		(a) $E_1$ & (b) $E_2$  & (c) $M$ \\
	\end{tabular}
	\caption{\textbf{Performance of PrivRec with Different FL Parameters.}}
	\label{fig:exp:flparm}
\end{figure*}

\subsection{Recommendantion Performance of PrivRec (RQ1)}
To answer RQ1, we implement PrivRec without DP according to Algorithm 1. For its DNN structure, the embedding dimension for each category feature is $64$, the number of hidden layers is $4$, and the nonlinear activation is the Sigmoid function. For other hyper-parameters, we adopt the optimal hyper-parameters following the study of RQ2. For  Movielens dataset, we set $E_1$, $E_2$ and $M$ to be 80, 100 and 20 respectively. For Frappe dataset, we set $E_1$, $E_2$ and $M$ to be  40, 100 and 30 respectively.

Before we proceed, we validate the correctness of PrivRec in the FL environment.
We compare the performance of PrivRec and PrivRec-CEN, where PrivRec-CEN means running the same neural network model in the centralized environment.
For PrivRec-CEN, we simply run it in the centralized multi-task setting where we treat each user as a task.
Also, we compare PrivRec-CEN with a state-of-the-art deep learning-based recommender system  Neural Collaborative Filtering (NCF)~\cite{he2017neuralCF}, which could considered as the upper bound of recommendation performance in our evaluation.
Their results are shown in  Table~\ref{tb:overall1} and Table~\ref{tb:overall2} for the  Movielens and Frappe datasets respectively.
On both datasets, there exists a slight performance degradation for PrivRec compared with PrivRec-CEN, but this is natural and inevitable due to the distribution design and how the gradients are merged for global model updates.
Overall, the performance of PrivRec is comparable to that of PrivRec-CEN, hence we are confident that our PrivRec model produces almost identical results as the centralized setting. We also observe that NCF achieves the best results among all methods in centralized training environment. The main reason is that NCF can fully utilize the user identifications to model the user-item interactions, however, in our PrivRec, the user identifications are excluded due to privacy concerns and its distributed training purpose.

As the core of PrivRec is its capability of performing recommendation in an FL environment, we compare PrivRec with the following state-of-the-art FL recommender systems that are most relevant to ours:

\textbf{FedGNN}~\cite{wu2021fedgnn}. FedGNN is a federated recommender system based on graph neural network (GNN), which employs local differential privacy to enhance user's sensitive data and adopts homomorphic encryption to encrypt interacted items for information exchange among clients.  FedGNN achieves the state-of-the-art performance in FL recommender system.

\textbf{NN-SELF}~\cite{chen2018federated}. In this work, the authors propose a DNN-based recommender system running in the FL environment, and it employs a  two-stage meta-learning scheme to achieve fast on-device personalization. We adopt its \texttt{SELF} setting where the training set is the support set of the corresponding testing user. The rest setting is identical to ours, including the input features and neural network structures.

\textbf{FOFL-SELF}~\cite{jalalirad2019simple}. This FL recommender system is proposed to address the flaws of second-order meta-learning by employing REPTILE to implement the first-order meta-learning. This is also a DNN-based model but does not utilize the side information of users and items.

\textbf{FCF}~\cite{ammad2019federated}. Federated Collaborative Filtering (FCF) is an FL-based implementation of matrix factorization. The authors formulate the updating rules to update the model parameters to suit the FL setting. 

The comparison results are also shown in Table~\ref{tb:overall1} and Table~\ref{tb:overall2}  for the  Movielens and Frappe datasets respectively. From the results, we can draw the following observations:
(1) FOFL-SELF and PrivRec outperform NN-SELF, demonstrating that the first-order REPTILE meta-learning is more capable of handling the scarce data of user clients than MAML that requires the local dataset to be split into support and query sets for training, and this is especially crucial for building an FL recommender system where a large portion of users are inactive.
(2) FedGNN, PrivRec, FOFL-SELF and NN-SELF have higher accuracy than FCF. This is because they adopt deep neural networks as their main building blocks, which have larger learning capacities in capturing user preferences from raw data.
(3) FedGNN achieves significant improvement compared to other baseline methods, which is largely contributed to the power of the graph embedding approach that can capture more complex relations between users and items. 
(3) Our PrivRec consistently outperforms all baseline FL methods. In particular, PrivRec outperforms FOFL-SELF. One of the main reasons is that FOFL-SELF does not utilize the side information of users and items, weakening its ability of addressing the data sparsity problem. In contrast, PrivRec fully takes advantage of the user/item features for representation learning. 
PrivRec also outperforms FedGNN, because FedGNN exploits encrypted techniques such as homomorphic encryption for information exchange and construct the graph, which damages the recommendation performance. 

\subsection{Impact of FL Hyperparameters (RQ2)}
To answer RQ2, we vary several important FL hyper-parameters of PrivRec and report the corresponding \textit{Hits@20} on both datasets.
These hyperparameters include the number of training epochs on the server and clients $E_1$ and $E_2$, and the number of clients in each training batch $M$.
Specifically, we vary the values of $E_1$, $E_2$ and $M$ in $\{5, 20, 40, 60, 80, \\ 100\}$, $\{5, 20, 40, 60, 80, 100, 120\}$ and $\{2, 5, 10, 15, 20, 25,\\ 30, 35, 40\}$ respectively.

The experimental results are illustrated in Figure~\ref{fig:exp:flparm}.
For $E_1$, the model performance increases dramatically when $E_1$ is smaller than 40, especially on Frappe dataset. When $E_1$ is over 40, the recommendation performance tends to remain stable on both datasets. For $E_2$, the model performance gradually increases, and remains stable when it reaches 100. This demonstrates that, our meta-learning based approach is able to achieve on-device personalization with sufficient iterations of local updates, yielding high recommendation accuracy at both the individual and system level. At the same time, varying $M$ shows the smallest impact to model performance, but the trend shows PrivRec prefers a relatively larger value of $M$. Overall, it is ideal to set $M>25$. 

These empirical observations and conclusions can guide us to assign the optical values to the model training in practice.

\begin{figure}[!b]
	\centering
	\includegraphics[scale = 0.51]{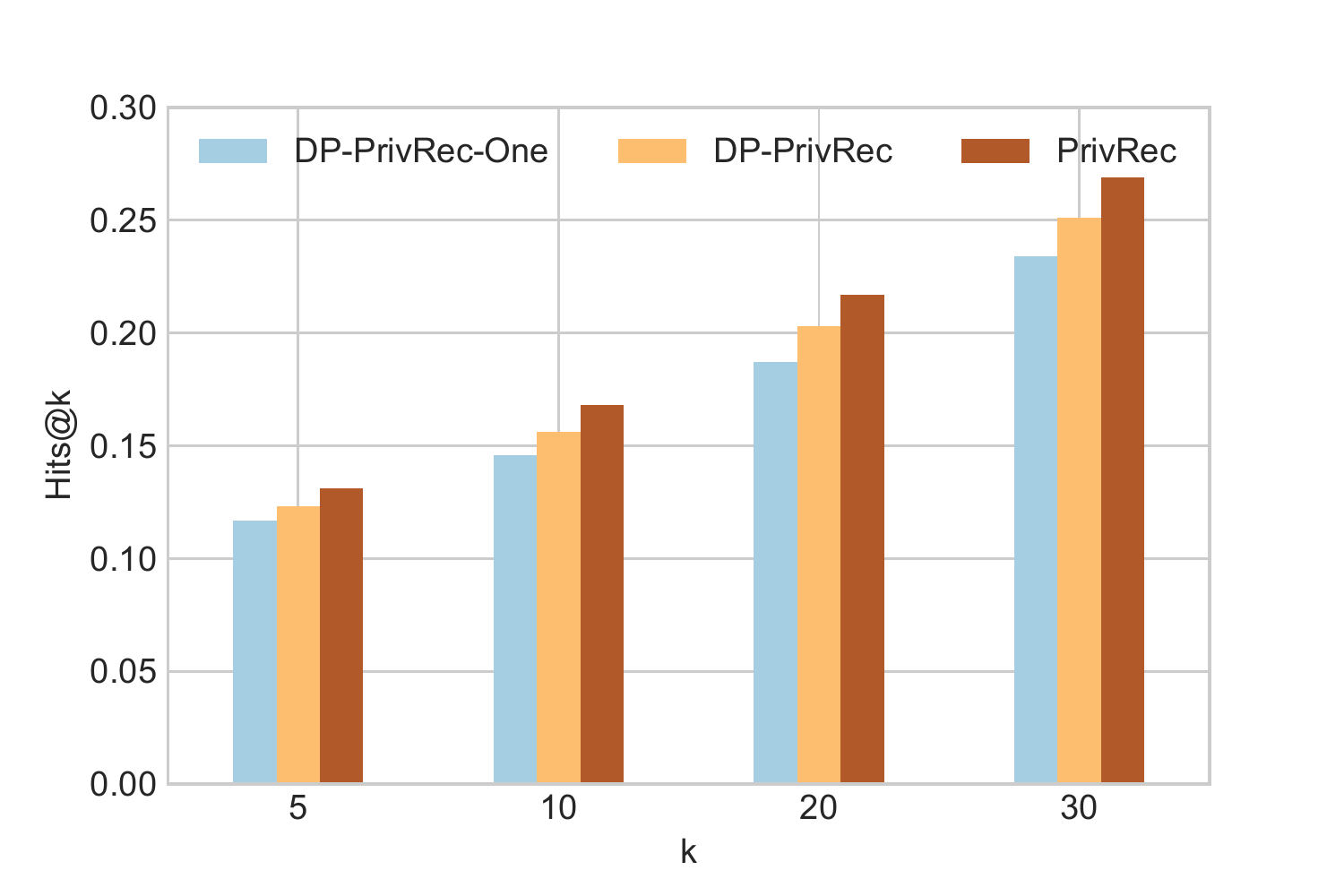}
	\caption{\textbf{\textit{Hits@k} of DP-PrivRec-One, DP-PrivRec and PrivRec on Movielens.}}
	\label{fig:exp:dpprivrec1}
\end{figure}

\begin{figure}[!b]
	\centering
	\includegraphics[scale = 0.51]{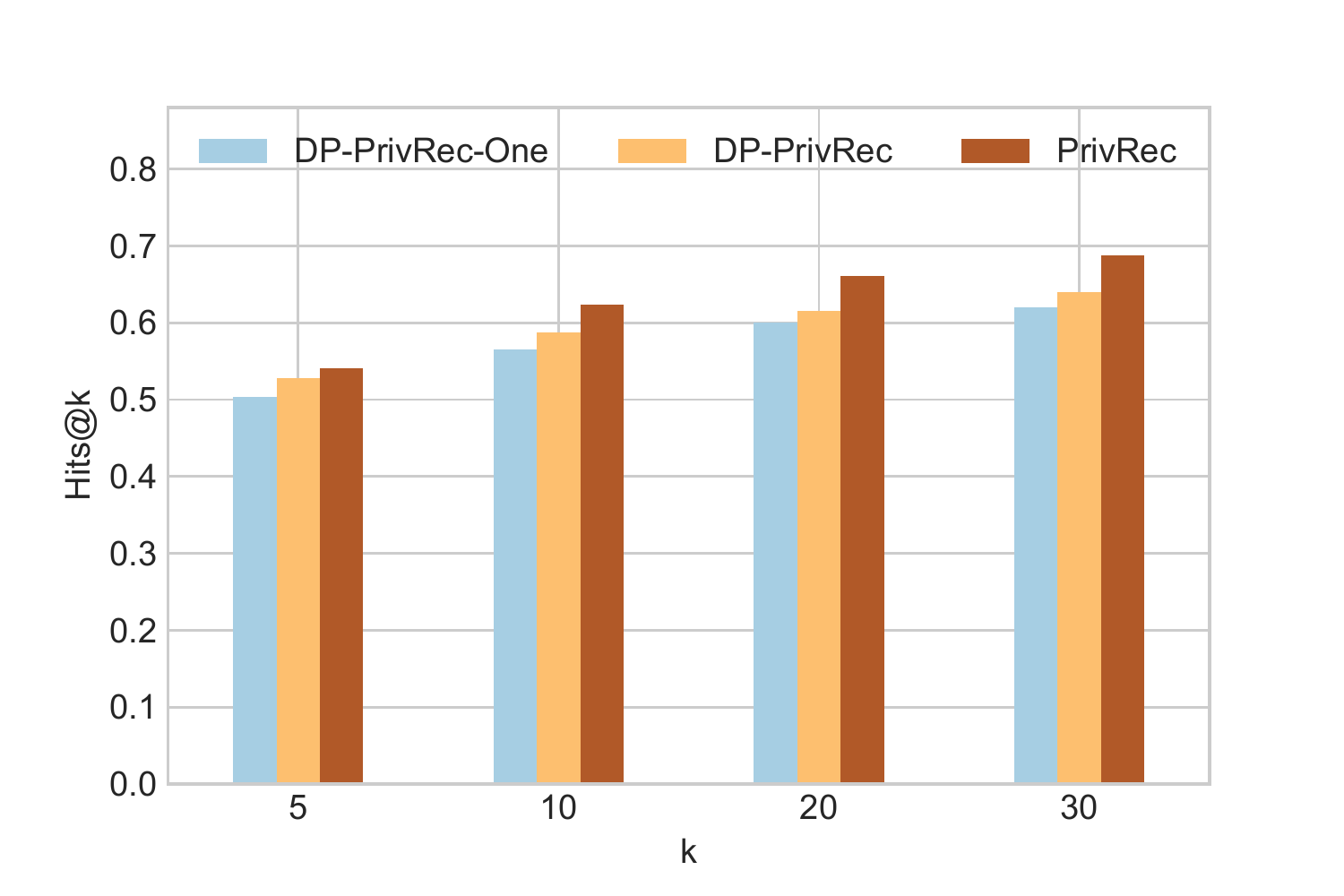}
	\caption{\textbf{\textit{Hits@k} of DP-PrivRec-One, DP-PrivRec and PrivRec on Frappe.}}
	\label{fig:exp:dpprivrec2}
\end{figure}

\subsection{Influence of Differential Privacy (RQ3)}
To answer RQ3, we run DP-PrivRec and PrivRec on both datasets, and compare their performances (\textit{Hits@k}). Recall that the main difference between them is that DP-PrivRec adds  Gaussian noises to the gradients used to update the global model to defend from adversary. For the hyper-parameters of DP-PrivRec in this experiment, we use $M=30$, and the bound $S$ is set as 40 for Movielens and 35 for Frappe respectively.
We also develop a variant of DP-PrivRec called \textbf{DP-PrivRec-One} by removing the SSL item representation learning component in the first stage, namely, it is optimized  based on only 
$\mathcal{L}_\mathrm{DSSM}$.

We show the results in Figure~\ref{fig:exp:dpprivrec1} and Figure~\ref{fig:exp:dpprivrec2} for both datasets.
From these figures, we can come to two major conclusions.
First, DP-PrivRec has lower prediction accuracy than PrivRec, which is within our expectation because we add a noise in the gradients during training to achieve user-level DP, making the model update less effective.
Second, DP-PrivRec exhibits comparable performance with PrivRec. Moreover, the performance gap between them is within 10\%, which is a tolerable trade off considering the much stronger privacy protection that DP-PrivRec offers. Third, DP-PrivRec consistently outperforms DP-PrivRec-One, validating our proposed two-stage training approach to effectively compensate the performance loss caused by introducing Gaussian noises into the model.

\begin{table}
	\center
	\caption{\textbf{$\epsilon$, Given Different $\delta$ and $q$ on Movielens. }}
	\begin{tabular}{|c|c|c|c|c|}
		\hline
		\backslashbox[10mm]{$q$}{$\delta$} & 0 &  1e-8 &   1e-6 &  1e-4 \\
		\hline
		5 / 4800 & $\infty$ &1.7439 & 1.3602 &  0.9764 \\
		\hline
		10 / 4800 & $\infty$  & 3.2117&2.7511 & 2.2535\\
		\hline
		15 / 4800 &  $\infty$ & 5.9705& 5.2030&4.3074 \\
		\hline
		20 / 4800 &  $\infty$ & 9.4736&8.3223 &6.9235 \\
		\hline
		25 / 4800 &  $\infty$ &13.7047  & 12.1696 & 10.2052 \\
		\hline
		30 / 4800 & $\infty$ & 18.9107 & 16.6081 & 14.3056 \\
		\hline
	\end{tabular}
	\label{tb:epsilon1}
\end{table}

\begin{table}
	\center
	\caption{\textbf{$\epsilon$, Given Different $\delta$ and $q$ on Frappe. }}
	\begin{tabular}{|c|c|c|c|c|}
		\hline
		\backslashbox[10mm]{$q$}{$\delta$} & 0 &  1e-8 &   1e-6 &  1e-4 \\
		\hline
		2 / 760 & $\infty$ & 2.2994  & 1.8388 & 1.3783 \\
		\hline
		5 / 760 & $\infty$  & 7.4618 & 6.5408 & 5.3932 \\
		\hline
		8 / 760 &  $\infty$ & 16.2853 & 14.3169 & 12.0143  \\
		\hline
		10 / 760 &  $\infty$ & 23.7651 & 21.4626 & 18.6182 \\
		\hline
		12 / 760 &  $\infty$ & 34.5025 & 30.0690 & 25.4638 \\
		\hline
		15 / 760 &  $\infty$ & 50.1537 & 45.5485 &  40.9433 \\
		\hline
	\end{tabular}
	\label{tb:epsilon2}
\end{table}

\subsection{Effects of Privacy Protection (RQ4)}
In this study, we vary the key DP parameters in DP-PrivRec and investigate the corresponding privacy loss $\epsilon$ and model performance.

\subsubsection{Privacy Loss.} First, we compute the upper bound of the privacy loss $\epsilon$ in $(\epsilon, \delta)$-DP using the RDP-based analytical Moment Accountant,  when given different values of $\delta$, the sampling ratio $q=M/N$ and the number of mechanism composition $\ensuremath{E}_1$. Note that we measure the privacy loss during the model training stage, meaning that $N$ is  80\% of the total users (i.e., 4800 and 760 in Movielens and Frappe respectively). We fix the noise variance $z=1$ for Gaussian Mechanism following previous work~\cite{thakkar2019differentially,mcmahan2017learning}, and set $\ensuremath{E}_1 = 1000$ for both datasets. We show the results for Movielens and Frappe in Table~\ref{tb:epsilon1} and Table~\ref{tb:epsilon2} respectively. Clearly, we can achieve different levels of privacy as needed by choosing different combinations of parameters. 
In particular, $\epsilon$ obtained on Movielens is clearly less than that on Frappe. This implies that using smaller sampling probability yields a tighter privacy budget, which is consistent with the privacy amplification rule~\cite{kasiviswanathan2011can}.
Therefore, for Frappe which has a smaller $N$, we need a smaller batch size ($M$) to achieve a similar level of privacy as on Movielens.


\begin{figure}[h]
	\centering
	\includegraphics[scale = 0.45]{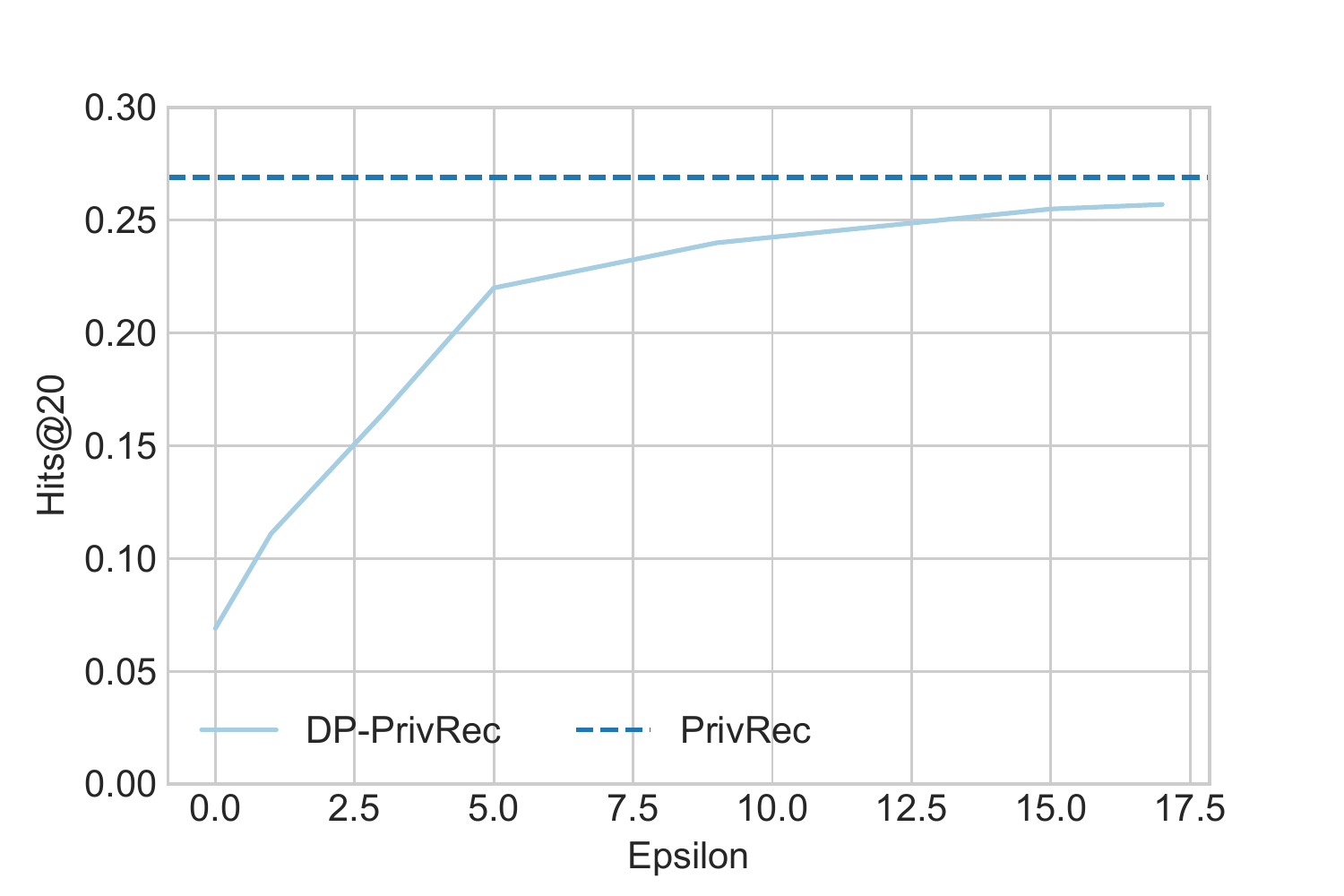}
	\caption{\textbf{\textit{Hits@20} of PrivRec and DP-PrivRec under different privacy budget $\epsilon$ on Movielens.}}
	\label{fig:budgetml}
\end{figure}
\begin{figure}[h]
	\centering
	\includegraphics[scale = 0.45]{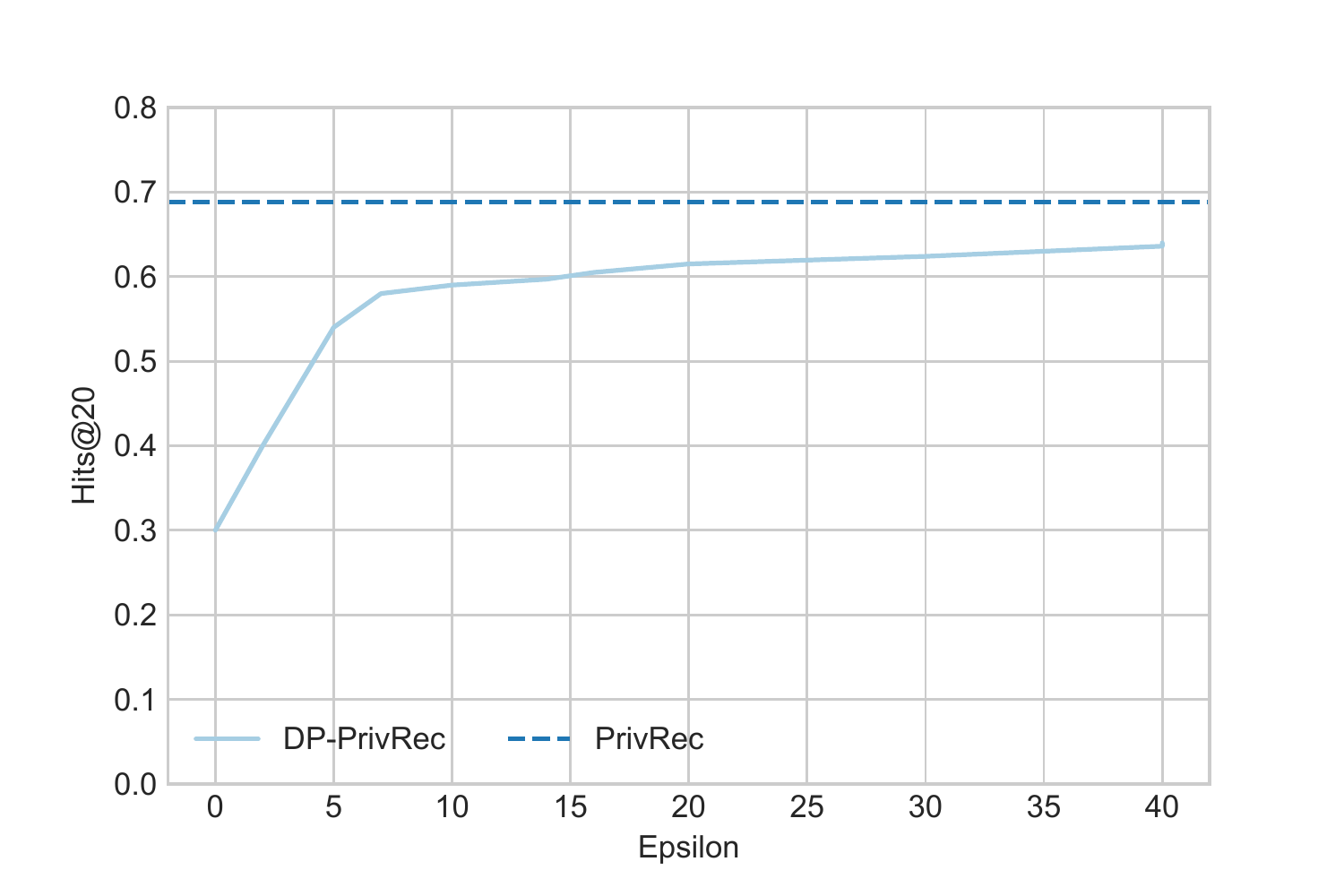}
	\caption{\textbf{\textit{Hits@20} of PrivRec and DP-PrivRec  under different privacy budget $\epsilon$ on Frappe.}}
	\label{fig:budgetfr}	
\end{figure}

\subsubsection{Impact of Privacy Budget on Accuracy}
In this section, we quantitatively investigate the impact  of privacy budget on recommendation accuracy.
Specifically, we vary the value of $\epsilon$ and observe the recommendation accuracy (\textit{Hits@20}) of DP-PrivRec on Movielens and  Frappe, we also show the performance of PrivRec in the same results for comparison.

We report the results in Fig. ~\ref{fig:budgetml} and Fig.~\ref{fig:budgetfr}, from which we can draw two conclusions.
First, PrivRec  can serve as an upper bound for DP-PrivRec in prediction accuracy, since there is no disturbance in model parameters. Second, the relation between the privacy budget and the target accuracy can be fitted as a log-like curve.  Increasing the  privacy budget (i.e., $\epsilon$) can quickly increase the prediction accuracy. However, as $\epsilon$ is sufficiently large, the accuracy remains stable.

\subsection{Effects of First-order Meta-learning (RQ5)}
To answer RQ5, we develop two experiments. The first one is to study the effects of the proposed meta-learning technique in helping the global model personalize to inactive users.  The second one is to compare the efficiency of first-order and second-order meta-learning in PrivRec.

\subsubsection{Effects of Meta-learning in Fast Personalization}
\begin{figure}[!t]
	\centering
	\includegraphics[scale = 0.45]{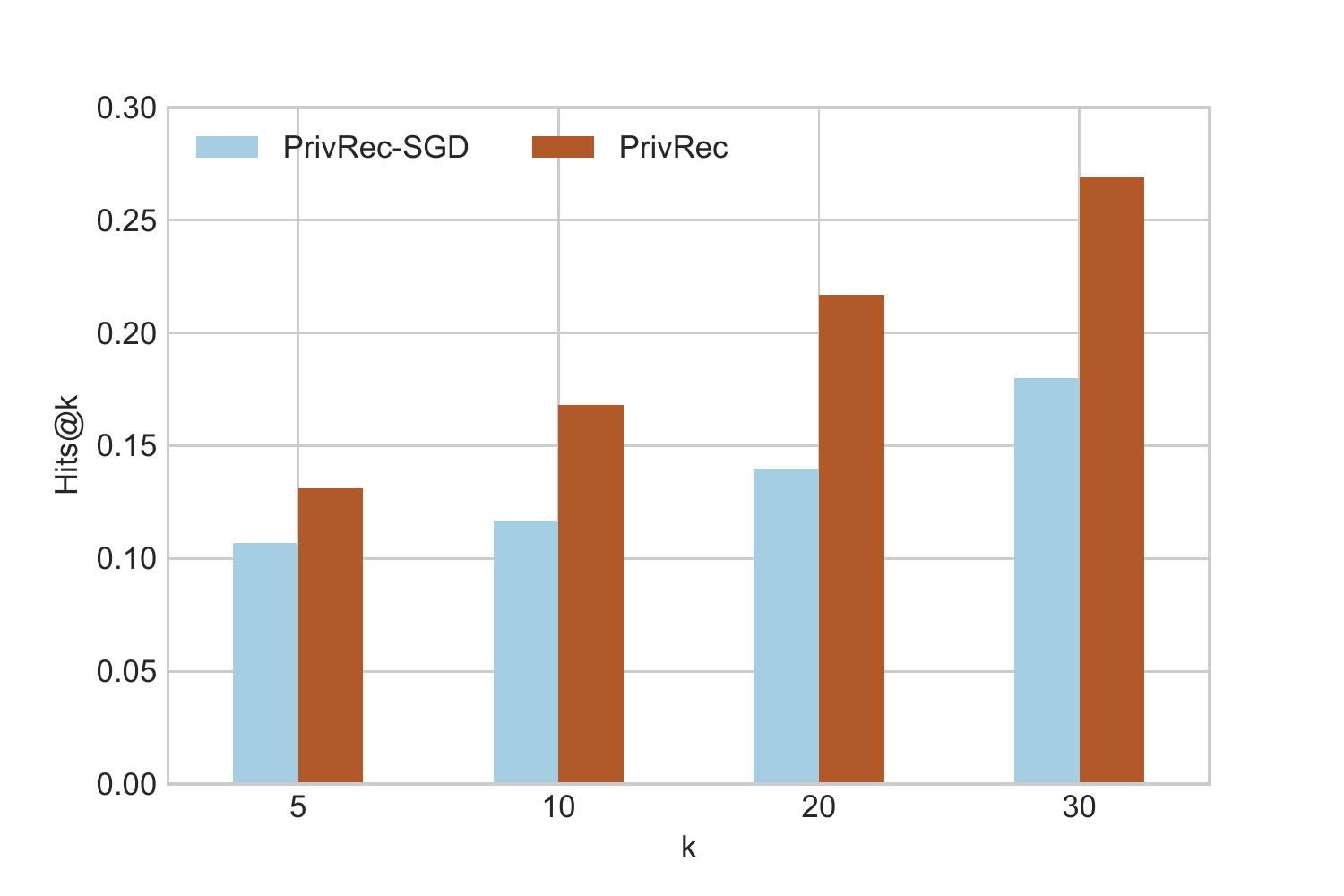}
	\caption{\textbf{\textit{Hits@20} of PrivRec-SGD and the full PrivRec on Movielens.}}
	\label{fig:PrivRec-SGD}
\end{figure}

We develop a variant of PrivRec called \textbf{PrivRec-SGD}, which only performs one gradient descent in the device during  training and  local personalization. This setting degenerates to SGD.
We compare it with PrivRec that utilizes a first-order gradient performing multiple local updates, and choose \textit{Hits@20} as the measurement. 
In Movielens, we select the inactive users who have less than 20 visited items in the test set for evaluation. 
We show the results in Fig.~\ref{fig:PrivRec-SGD}. From the figure, we can see that PrivRec significantly outperforms PrivRec-SGD by   $20\% - 40\%$.
The huge gap between these two methods demonstrates that meta-learning can be especially beneficial to those inactive users.

\subsubsection{Efficiency of First-order Meta-learning}
\begin{figure}[!t]
	\centering
	\includegraphics[scale = 0.45]{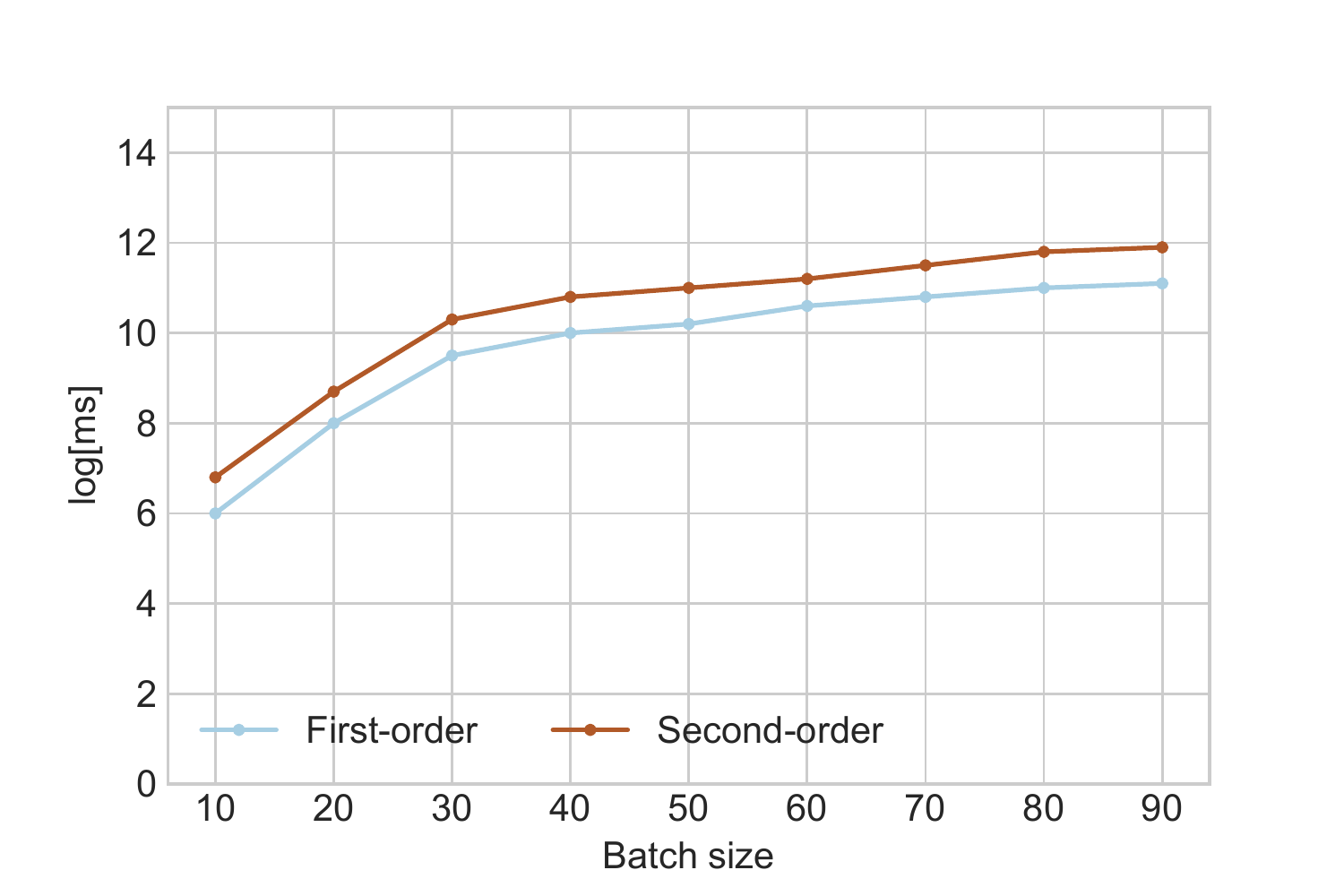}
	\caption{\textbf{The time  spent (in log of milliseconds) in every training round for FL recommendation based on first- and second-order meta learning.}}
	\label{fig:efficiency}
\end{figure}

We compare the training time of our proposed PrivRec that  uses  first-order meta-learning and the baseline method NN-SELF~\cite{chen2018federated}  that  uses second-order meta-learning (i.e., MAML). Note that  we employ the same network structure  as NN-SELF for the base recommender system in this experiment for fair comparison.
Specifically, we vary the batch size (i.e., the number of participating users) in every global training round on Movielens, and compare the time cost (log of milliseconds) by these two methods in Fig.~\ref{fig:efficiency}. 

From the results, the observation agrees with the theoretical study, such as~\cite{nichol2018first,finn2017model,rajeswaran2019meta}, that computing the expensive second-order Hessian-vector products in an additional backward pass could be about $10\% - 30\%$ more time consuming. This justifies the need to develop a first-order meta-learning approach.

\subsection{Defensing Membership Inference (RQ6)}
To answer RQ6, we develop a membership inference attack (MIA), and evaluate the effects of DP-PrivRec against the attacks.

Following previous studies~\cite{truex2019effects,rahman2018membership}, we design a shadow model $f_{s}$ and an MIA model $f_{mi}$, where  $f_{s}$ is used  to generate a dataset for training $f_{mi}$.
Assume that there is a malicious FL participant, when it receives the global model without adding noise, the MIA model $f_{mi}$ in this participant determines whether a specific given user is present in the FL training or not.

\subsubsection{Training Attack Model}

We denote  PrivRec as the target model, and let the shadow model $f_{s}$ trained on a \textit{shadow dataset} generated by the adversary to mimic the target model, namely, take similar input and output of the target model. 
This shadow dataset should be created by publicly available data and have the same distribution (e.g., features) as the original one.
Specifically, in our experiment, we randomly select 1000 users from Movielens introduced in Section~\ref{st:dataset} and their interactions to form shadow dataset for simulation purpose. In reality, we can consider that some users are willing to sacrifice their privacy and turn out their data in exchange of better personalized service. We call the remaining dataset excluding the shadow dataset as private dataset.
We randomly select $80\%$ users and their histories from shadow dataset (denoted as used users) to train the shadow model  $f_{s}$, which has the same network structures and hyper-parameters as PrivRec. This requirement can be met since all participants are aware of the model structure and hyper-parameter information in FL.

After the training of $f_{s}$ completes, we construct an attack dataset based on the  $80\%$ used users, the rest   $20\%$ unused users and the shadow model.
We feed each user $u$ in shadow dataset into $f_{s}$, and obtain a  top-$10$ recommendation list $l_u$.
Since MIA is a binary classification task, namely, predicting whether a sample has involved in the training or not, we  add a label to such $(u, l_u)$ pair to indicate whether user $u$ is involved in training or not.
Specifically, if $u$ is from ``used users'', then we label $(u, l_u)$ as \textit{in}, if $u$ is from ``unused users'', then we label $(u, l_u)$ as \textit{out}.
We  concatenate $(u, l_u)$ as input, and employ random forest as the MIA model  $f_{mi}$ to predict the binary label \textit{in} or \textit{out}.

\subsubsection{Evaluating Defense Against MIA}

With well trained attack model  $f_{mi}$ and shadow model $f_{s}$, we evaluate the attack effectiveness of $f_{mi}$ and to what extend can DP-PrivRec defense such attacks. 
We divide the private dataset by user, and use $50\%$ of users to separately train PrivRec and DP-PrivRec. 
In one of the participating client, which is assumed to be malicious, a MIA model  $f_{mi}$ is trained following the aforementioned steps.
If it is curious whether a specific user represented by $u$ (i.e., a user has specific features) has participated in the training, PrivRec or DP-PrivRec that this participant receives could predict the top-$10$ recommendation items $l_u$. Given $(u, l_u)$, $f_{mi}$ predicts the label  \textit{in} or \textit{out}. We compare the prediction accuracy of MIA $f_{mi}$ based on all output of PrivRec or DP-PrivRec.
Furthermore, to justify the advantage of Gaussian Mechanism in adding noise to defense from membership inference, we develop a variant \textbf{LM-PrivRec} that draws noise from  Laplace Mechanism, another popular DP mechanism,  to defense from membership inference attack. The settings of Laplacian distribution from which noises are drawn from follow prior work~\cite{uhlerop2013privacy}.

We show  results of this experiment in Table~\ref{tb:attack}. We can come to four major conclusions based on these results. First, the parameters of PrivRec are not perturbed by any noises, therefore, when the adversary  launches MIA to its results, the prediction accuracy is the highest among all target models. Second, the  prediction accuracy of the attack model to LM-PrivRec and DP-PrivRec is  lower than that of  PrivRec, which indicates that  introducing DP to PrivRec to generate noisy results could effectively defend from MIA. Third, by setting higher privacy budgets ($\epsilon$) in DP-PrivRec, MIA will achieve better performance, which also usually means more sacrifice on privacy and better recommendation results. Finally, LM-PrivRec has slightly worse protection effectiveness than DP-PrivRec given the same $\epsilon$ towards MIA because Gaussian Mechanism could introduce less noise into the model while having the same privacy budget.

\begin{table}[h]

	\centering
	\caption{\textbf{The accuracy of membership attacks to different target models.}}
	\begin{tabular}{|l|l|l|}

		\hline
		\backslashbox[36mm]{Target Model}{Dataset} & Movielens  &  Frappe \\  \hline
		PrivRec                                    & \textbf{0.65}      & \textbf{0.73}   \\ \hline \hline
		DP-PrivRec ($\epsilon=2$)                   & 0.51      & N/A    \\ \hline
		DP-PrivRec ($\epsilon=5$)                   & 0.54       & N/A   \\ \hline
		DP-PrivRec ($\epsilon=15$)                   & 0.55       & N/A   \\ \hline LM-PrivRec ($\epsilon=15$)                  & 0.59      & N/A    \\ \hline \hline
		DP-PrivRec ($\epsilon=5$)                   & N/A       & 0.54  \\ \hline		
		DP-PrivRec ($\epsilon=20$)                  & N/A      &   0.61   \\ \hline
		DP-PrivRec ($\epsilon=40$)                  & N/A       & 0.65   \\ \hline
		LM-PrivRec ($\epsilon=40$)                 &  N/A      & 0.69    \\ \hline
	\end{tabular}
\label{tb:attack}
\end{table}

\section{Related Work}
\label{st:related}
In this section, we review recent advances relevant to our work in three lines: federated learning, personalized FL algorithms and private FL algorithms.
\subsection{Federated Learning}
With the boom of machine learning, especially deep neural networks (DNN) over the past decade, numerous practical applications based on these techniques have emerged to help users address real-world problems, such as recommender system~\cite{chen2019air,zhang2019deep,zhang2018discrete,sun2020go,zhang2020gcn}, facial recognition~\cite{masi2018deep}, AI assistance~\cite{wang2019origin} and argument mining~\cite{hung2017computing,nguyen2017argument}.
These models  usually run in a cloud-based paradigm, namely these machine learning models are trained and hosted on cloud servers, and they provide on-demand services for users~\cite{chen2021learning,wang2020next}.
However, DNN-based models are notoriously known to be data-hungry, and thus require access to a huge amount of user data.
Meanwhile, many countries enforce  laws and regulations to protect the personal information privacy and security.
These laws and regulations make online service providers hard to collect and centrally store user data for training purposes.

Federated learning (FL) is a new attempt to solve the data dilemma faced by traditional machine learning methods, which enables training a shared global model with a central server while keeping all the sensitive data in local institutions where the data belong.
Google was the first to propose federated learning concept in 2016, and applied this technology to their application Gboard - a virtual keyboard of Google for touchscreen mobile devices with support for more than 600 language varieties~\cite{mcmahan2017learning,hard2018federated,ramaswamy2019federated}.
Since then, various machine learning algorithms have been proposed  to adapt to the FL setting. Based on how the data across parties are utilized in FL, these algorithms can be categorized into three categories according to~\cite{yang2019federated}: horizontal federated learning such as the aforementioned Gboard,  vertical federated learning such as~\cite{hardy2017private,nock2018entity} and federated transfer learning.

The exploration of building an FL recommender system, which falls into the horizontal federated learning category, is still rare.
There are some work aiming to develop an FL matrix factorization recommender system ~\cite{flanagan2020federated,ammad2019federated}.
~\cite{ammad2019federated} introduce a Federated Collaborative Filter (FCF) model that generates recommendation results based on implicit feedback data by deriving a federated version of the widely used  CF method. To update the global model, FCF aggregates user-specific gradient updates of the model weights from the clients.
In ~\cite{flanagan2020federated}, the authors propose a federated multi-view matrix factorization method for recommendation, which enhances the performance by including  side information from both users and items.

We are the first few work to propose a DNN-based personalized FL recommender system that fully utilizes the user/item side information to learn their representations.

\subsection{Personalizing FL algorithms}
On the other hand, a typical solution to obtain an optimal FL model is only good on average, and it does not fully consider the heterogeneity of data distribution of users~\cite{fallah2020personalized}.
This contradicts our goal to develop a practical recommender system that can generate personalized results for each user. 
Therefore, it is necessary to ``personalize'' the vanilla FL algorithm.
Currently, there are three major lines of work addressing this challenge: local fine-tuning, multi-task learning and adding user context.

\textbf{Local fine-tuning} The mainstream approach to personalize an FL algorithm is local fine-tuning, where each client receives 
a global model and tunes it using their own local data and several gradient descent steps.
For example, ~\cite{deng2020adaptive} propose to adaptive personalize federated learning   algorithm which aims to learn a personalized model for each user that is a mixture of optimal local and global models.
Recently, this approach is predominantly used in meta-learning methods such as MAML~\cite{finn2017model}.
In~\cite{jiang2019improving}, a personalized FedAvg algorithm in which the classic FedAvg algorithm is first deployed, and then they switch to REPTILE, a meta-learning algorithm proposed in~\cite{nichol2018first}, and finally runs local updates to achieve personalization. They found that FL with a single objective of performance of the global model could limit the capacity of the learned model for personalization.
ARUBA proposed in~\cite{khodak2019adaptive} is a meta-learning algorithm inspired by online convex optimization, and it achieves improved personalization when applied to  FedAvg.
The most closely related work to ours is NN-SELF~\cite{chen2018federated} where the authors build an FL recommender system based on MAML, where a parameterized meta-algorithm is used to train parameterized recommendation models and both meta-algorithm and local model parameters need to be optimized. 
Our work is different from NN-SELF in two aspects.
First, our model is a  deep learning-based structure that is specifically designed for FL environment. In contrast, the base recommender in NN-SELF  hinders its ability to learn user and item representations more accurately. Second, we have fully justified our motivation of replacing MAML with REPTILE, and the rationale why it leads to significant benefits. As we discussed, there is no need to split the local into support and query sets, which is more friendly for inactive users. Also, it can also effectively reduce the computational cost.

\textbf{Multi-task learning} The second category of such work is to view the personalization problem as  multi-task learning~\cite{zhang2017survey,caruana1997multitask}. 
The most well-known work in this category is MOCHA~\cite{smith2017federated}, which considers the optimization on each client  as a new task such that the approaches of multi-task learning can be applied.
In another work~\cite{mansour2020three}, the authors propose to cluster clients into groups based on some features and consider each of them as similar tasks, and then train a model for each group.
However, in such setting, all clients are required to participate in every training round, which is infeasible for a large-scale federated learning system.

\textbf{Adding user context} Finally, we introduce the third category that the personalization of a global FL model could be achieved by adapting to different contexts. For example, Hard \textit{et al.} in~\cite{hard2018federated} develop an FL next-word prediction model in a virtual keyboard for smartphones and personalize it on the local device by incorporating different contexts.

Our work falls into the first category, but different from their work, our proposed (DP-)PrivRec only utilizes the more efficient first-order gradient to adapt to the resource-constraint edge devices, while achieving great on-device personalization.

Also, the heterogeneity problem in FL has drawn much research attention, such as~\cite{zhao2018federated,wang2020optimizing,briggs2020federated,xu2021federated}. 
Compared with the non-IID concept discussed in~\cite{zhao2018federated,wang2020optimizing,briggs2020federated}, our work is more focused on the unbalanced nature of activities users generate, namely, the number of activities each user (device) varies significantly, which greatly challenges the personalization requirement of recommender systems. However, these mainly studies the distribution differences among devices, which is more concerned on the global optimization target. 

\subsection{Private FL Algorithms}

In a pure FL environment, the clients do not need to transfer data to the third party during model training. However, transferring parameters or gradients may still be susceptible to leaking sensitive information~\cite{melis2019exploiting,wang2019adaptive,wang2019beyond}. 
Membership inference attack (MIA) is considered as a major threat in such circumstance, which aims to detect if some individual's data has been used to estimate a statistical model.
Extensive techniques have been made to defend from MIA during aggregating model parameters, and the most well-studies ones are Secure Multiparty Computation (SMC)~\cite{evans2017pragmatic,bonawitz2017practical}, Homomorphic Encryption (HE)~\cite{acar2018survey}, and Differential Privacy (DP)~\cite{dwork2014algorithmic}.
With HE, a client and server can exchange information and perform training based on encrypted data. HE is quite computation intensive and inflexible, so it has not been widely used in the production environment.
SMC designs a network consisting of multiple carefully selected parties, who collaboratively carry out a given computation task. Each party in the network has access to only an encrypted part of the data. Some simple functions have been implemented based on SMC, however, arbitrarily complex function computations such as deep learning are still infeasible due to computational cost~\cite{makri2019epic}.
DP aims to reduce the statistical likelihood of any individual data sample being identified as a training sample.

Compared with SMC and HE, DP has its unique advantages when applied in the FL setting. First, it provides strong theoretical privacy guarantees without any information about an individual and is not affected by what the attackers know about the dataset. The benefit is that the involved parties can access datasets without ad-hoc restrictions, maintaining the advantages of the  FL training  framework.  Second, a very nice feature of DP is that it can be composed, which we also cover in our submitted manuscript. Specifically, in modern machine learning training process, many times of access to training data are required, which would naturally result in more data exposure. Theoretically, the composition feature of DP enables it to achieve minimized privacy loss in various settings when accessing dataset many times. In our setting, we use the moments accountant mechanism~\cite{abadi2016deep} to keep track of and limit the privacy loss when training our proposed deep learning-based recommender system.

Next, we review the work~\cite{geyer2017differentially,mcmahan2017learning,hamm2016learning,zhang2019efficient,ribero2020federating} that apply DP in FL.
The common goal of these work is to ensure that a learned model does not reveal whether a client participated during decentralized training.
This requires the protection of the  client's whole dataset against differential attacks from other clients.

Specifically, the FedMEC framework in~\cite{zhang2019efficient} is an efficient federated learning service on the mobile edge computing environment, which allocates the heavy computations to the edge devices and makes the computation results differentially private before sending back to the server.
On the other hand, \cite{mcmahan2017learning} and \cite{geyer2017differentially} independently propose a user-level DP algorithm in the federated learning setting and provide a tight privacy guarantee. 
In~\cite{ribero2020federating}, the authors also propose a FL recommender with differential  privacy, which divides users into multiple groups, and separately learns a DP prototype recommendation model for each entity. However, this assumes that there is no privacy concerns for users within a group.
Most recently, ~\cite{li2020federated} introduce a DP-based  Upper Confidence Bound (UCB) strategy to the proposed federated private bandits framework in order to protect the client's data from exposure, and based on this scheme, the authors build a recommender system.

In our work, we formulate a user-level differentially private PrivRec called  DP-PrivRec. Compared to the DP techniques used in \cite{mcmahan2017learning,geyer2017differentially,ribero2020federating,li2020federated}, DP-PrivRec is different from them in two aspects.
First, we sample an equal-sized batch of clients instead of varied-size, which helps us easily analytically measure the privacy used during training and accelerate the computation. Second, we introduce a two-stage training approach to make up for the performance loss due to DP.

\section{Conclusions}
\label{st:conclude}
When widely used recommender systems meet the dramatically rising concerns of personal privacy protection in the mobile Internet era, it is urgent and practical to develop a recommender system that can balance privacy and recommendation performance. 

In this paper, we aim to address this dilemma.
We first proposed a DNN-based recommender system called PrivRec that utilizes the side information of users and items to learn the user/item representations, and then generate recommendations. In particular, PrivRec can smoothly run on the FL setting, which is a fully distributed framework to design privacy-persevering algorithms.
With PrivRec, sensitive user data never has to leave their devices, while still enjoying the recommendation service.
Since the trivial FL training paradigm can hardly provide personalized results for every individual user, we introduced an efficient method based on meta-learning to fast adapt to a new user or inactive user using a few local data examples. 
Furthermore, as there are still risks to leak personal information to potential malicious FL participants, 
we enhanced the privacy protection of PrivRec by introducing a user-level DP mechanism, and developed a more powerful model called DP-PrivRec. We also provided an analytical sketch to measure the privacy spent when training DP-PrivRec.
Finally, we performed extensive experiments on two datasets within a simulated FL environment, whose results demonstrate the effectiveness of our proposed PrivRec and DP-PrivRec.

\section{ACKNOWLEDGEMENT}
This work was supported by ARC Discovery Project (Grant No. DP190101985) and ARC Future Fellowship (FT210100624). 

\bibliographystyle{spmpsci}
\bibliography{sigproc}

\end{document}